\documentclass[conference]{IEEEtran}
\IEEEoverridecommandlockouts

\ifCLASSINFOpdf
\else
\fi
\usepackage{geometry}
\usepackage{adjustbox}
 \usepackage{mathptmx}      
\usepackage{xparse}%
\usepackage{multicol}
\usepackage{xcolor}
\usepackage{booktabs}
\usepackage{subfig}
\usepackage{xcolor,colortbl}
\usepackage[T1]{fontenc}
\usepackage{lipsum}
\usepackage[utf8]{inputenc}
\usepackage{array}
\usepackage{lettrine}
\usepackage{graphicx}
\usepackage{stfloats}
\usepackage{amsmath}
\usepackage{amssymb}
\usepackage{prettyref}
\usepackage{hyperref}
\usepackage{multicol}
\usepackage{subfig}
\usepackage[T1]{fontenc}
\usepackage{lipsum}
\usepackage[utf8]{inputenc}
\usepackage{array}
\usepackage{booktabs}
\usepackage{lscape}
\usepackage{authblk}
\usepackage{rotating}
\usepackage{stfloats}
\usepackage{xcolor}
\usepackage{rotating}
\usepackage{setspace}
\usepackage{multirow}
\usepackage{amsmath}
\usepackage{algorithm}
\usepackage{mathtools}
\usepackage{url}
\usepackage{algpseudocode}
\makeatletter
\usepackage[skip=2pt,font=footnotesize]{caption}

\begin{document}
	\newgeometry {top=25.4mm,left=19.1mm, right= 19.1mm,bottom =19.1mm}%
\title{
Fractional-order Modeling of the Arterial Compliance: An Alternative Surrogate Measure of the Arterial Stiffness
}
\author{Mohamed~A. Bahloul
        and Taous-Meriem Laleg Kirati
}
\markboth{Journal of \LaTeX\ Class Files,~Vol.~14, No.~8, August~2015}%
{Shell \MakeLowercase{\textit{et al.}}: Bare Demo of IEEEtran.cls for IEEE Journals}
\maketitle
\begin{abstract}
Recent studies have demonstrated the advantages of fractional-order calculus tools for probing the viscoelastic properties of collagenous tissue, characterizing the arterial blood flow and red cell membrane mechanics, and modeling the aortic valve cusp. In this article, we present novel lumped-parameter equivalent circuit models of the apparent arterial compliance using a fractional-order capacitor (FOC). FOC, which generalizes capacitors and resistors, displays a fractional-order behavior that can capture both elastic and viscous properties through a power-law formulation. The proposed framework describes the dynamic relationship between the blood pressure input and blood volume, using linear fractional-order differential equations. The results show that the proposed models present reasonable fit performance with in-silico data of more than 4,000 subjects. Additionally, strong correlations have been identified between the fractional-order parameter estimates and the central hemodynamic determinants as well as pulse wave velocity indexes. Therefore, fractional-order based paradigm of arterial compliance shows prominent potential as an alternative tool in the analysis of arterial stiffness.
\end{abstract}

\IEEEpeerreviewmaketitle

\section{Introduction}
\lettrine[findent=2pt]{{\textbf{O}}}{ver }the last decades, arterial models have been proven to be extremely useful and effective in unraveling cardiovascular diseases \cite{liang2005closed}, in the medical intervention planning \cite{alderliesten2004simulation}, in diseases' treatment and monitoring \cite{van2016patient}, and in the design and testing of medical devices and simulators \cite{beulen2011toward,huberts2018needed}. Besides, arterial models have shown great potential in the noninvasive evaluation of physiological parameters, which are not directly accessible, such as the arterial compliance and stiffness \cite{leguy2010estimation,stergiopulos1995evaluation,stergiopulos1999total}. Vascular compliance is defined as the ability of a particular arterial vessel to store blood. It describes the capacitance of the vascular wall to dynamically distend and increase the vessel volume with an increase in the transmural pressure or the tendency of the vascular wall to resist and recoil toward its original geometry with compression. Functionally, arterial compliance is demonstrated by the relationship between the stored blood volume's variation and the input blood pressure's variation. Similarly, the concept of total arterial compliance was introduced as the sum of all compliance components of the entire arterial system. Thus, the total compliance describes the global arterial capacity to store blood and is equal to the variation in blood volume in the entire arterial system divided by the systemic input pressure's variation. However, it is known that this ratio is not only governed by the total arterial compliance but also incorporates some other effects such as the pulse wave reflection. Indeed, it is equivalent to the total compliance only at low frequency. Hence, the concept of dynamic arterial compliance-or, equivalently, apparent compliance have been proposed by Quick $et \ al.$ \cite{quick1998apparent} to show how to estimate the true total compliance correctly from the transfer function relating the blood volume to the input pressure \cite{quick2000true} and explain a question of fact as to whether the classical estimation methods of arterial compliance fails to yield to true arterial compliance. Before the introduction of the "apparent compliance" concept, the transfer function relating blood volume to systemic input pressure is thought to be constant and modeled by a constant capacitance of an ideal capacitor (electrical analog model). This hypothesis is based on the Windkessel concept, which is adopted by the lumped-element modeling school. The drawbacks of this assumption are reflected in its estimation-based methods of compliance, which doesn't yield to a correct evaluation of the true arterial compliance \cite{craiem2003new}. Because of the distributed nature of the vascular compliance and resistance within the arterial network, the relationship blood Volume/input pressure is frequency-dependent \cite{quick1998apparent}. Accordingly, a time delay between the arterial blood volume and the input pressure occurs. During the past decades, some clinical studies demonstrated the necessity of introducing apparent compliance to extract total compliance. Therefore,  a new lumped-parameters modeling framework, which takes into account the complex and frequency-dependence properties of the apparent arterial compliance, have been proposed \cite{burattini1998complex}. These models are based on the idea that the arterial wall is viscoelastic rather than pure elastic. Hence, the Voigt cell model (resistor in series with a capacitor) has been proposed as a suitable candidate to represent the total arterial compliance. The resistor of the Voigt cell displays the viscous losses held by the arterial wall motions, while the capacitor represents the static compliance of the arteries. Through the combination of resistor and capacitor cells gives rise to complex and frequency-dependent compliance,  the Voigt model configuration is considered very poor in representing the arterial viscoelasticity properties since it does not account for the \textit{stress-relaxation} experiment \cite{visaria2005modeling}. Therefore, to address this inconsistency, the order of the viscoelastic representation has been increased by adding more viscous and elastic connected elements \cite{burattini1998complex}. The higher-order configuration provided a more accurate but complex configuration, where its complexity is principally due to the enormous number of unknown parameters, which suggests another challenge. Indeed, for higher-order models, the number of parameters to identify is more significant, while the collected real data is small and insufficient. It is also known that reduced-order models are desirable for their simplicity and ease of exploration.
Over the last decades, the fractional-order derivative (FD), defined as a generalization of the standard integer derivative to a non-integer order, has been gaining paramount popularity in modeling and characterizing biological tissues \cite{ionescu2010modeling,kobayashi2012modeling}. Because of its non-locality and memory properties, FD has been regarded as a powerful tool for modeling complex physical phenomena that exhibit \textit{power-law} response or involve memory effects \cite{magin2006fractional,jaishankar2013power}. In recent research, the power-law behavior has been proved in the viscoelasticity characterization of an elastic aorta. The \textit{in-vivo} and \textit{in-vitro} data analysis showed that the FD tools are more convenient to accurately model and describe the arterial wall viscoelastic dynamic response \cite{craiem2007fractional,craiem2008fractional,craiem2010fractional,perdikaris2014fractional,zerpa2015modeling}. Besides, a recent study by the authors \cite{bahloul2018three,bahloul2018arterial}, used fractional-order derivative tools to the well-known arterial Windkessel paradigm, by replacing the ideal capacitor, which accounts for the total arterial compliance, with a fractional-order capacitor. The preliminary analysis demonstrated that the fractional-order impedance is the right candidate for the accurate assessment of the aortic input impedance. Furthermore,  a simple correlation between the main parameters of the central arterial blood pressure and the fractional differentiation operator has been shown. Consequently,  the novel fractional-order parameter may have an influential role as a physiological index of the arterial stiffness \cite{bahloul2019fractional}. This paper introduces and investigates the fractional-order derivative modeling framework for apparent compliance. The proposed modeling framework offers a new paradigm for the physiological interpretation of the frequency-dependent arterial compliance and the interaction between the systemic arterial mechanical properties (viscosity and elasticity). Besides, this study compares the different proposed models as well as with the corresponding integer-order models. The rest of the paper is organized as follows: in Section II, the preliminaries, the proposed models and the method are presented. Section III shows the results and discussion. Finally, section IV presents the conclusion and future perspectives.
\section{Material and method}
\begin{figure*}[!t]
\vspace{-.5cm}
	\centering
	\includegraphics[height=5cm,width=17cm]{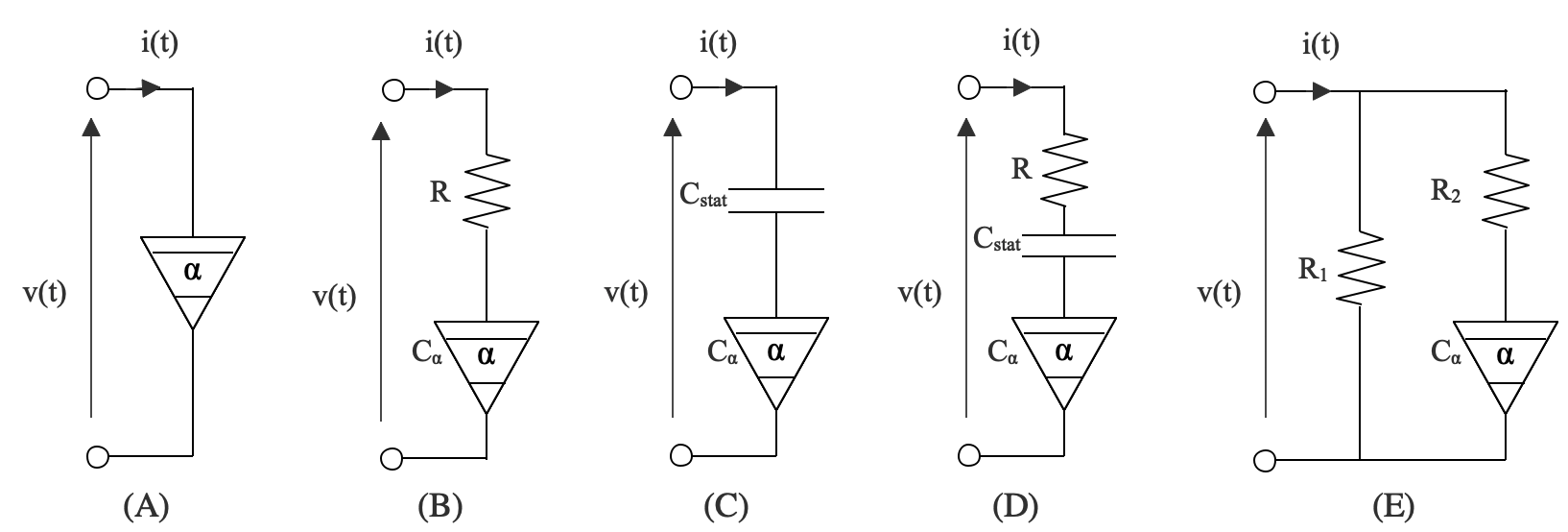}
	\caption{Schematic representations of the electrical analog of the proposed fractional-order models.}
	\vspace{-.5cm}
\end{figure*}
\subsection{ Preliminaries}
\subsubsection{Input impedance, apparent compliance, and resistance}
Aortic input impedance ($Z_{in}$) and apparent compliance ($C_{app}$) are considered significant in the characterization of the arterial system, independently of the heart properties. Whereas $Z_{in}$ describes the ability of the arterial system to hamper the blood flow dynamically, $C_{app}$ depicts the capacity of the arterial bed to store blood dynamically. Functionally, $Z_{in}$ is defined as the dynamic relationship, in the frequency domain, of the arterial blood pressure ($P_{in}$) and blood flow  ($Q_{in}$) at the entrance of the systemic circulatory system, that is:
\begin{equation}
   Z_{in}(\omega)=\dfrac{P_{in}(\omega)}{Q_{in}(\omega)},
\end{equation}
where $\omega$ corresponds to the angular frequency. 
$C_{app}$ is defined as the dynamic relationship, in the frequency domain, between the blood volume $V$, and the input aortic blood pressure ($P_{in}$) that is:
\begin{equation}
    C_{app}=\dfrac{V(\omega)}{P_{in}(\omega)}
\end{equation}
 Similarly, to the concept of apparent compliance, another frequency dependent transfer function relating $P_{in}$ to the output blood flow ($Q_{out}$) has been defined as well. It describes the so-called apparent resistance ($R_{app}$) and it can be formulated as:
\begin{equation}
    R_{app}=\dfrac{P_{in}(w)}{Q_{out}(w)}
\end{equation}
Based on Quik's et al. investigations in \cite{quick1998apparent} and \cite{quick2000true}, $R_{app}$ can be approximated as a constant that is equivalent to the total peripheral resistance. Additionally, $C_{app}$ can be expressed in terms of $Z_{in}$ and $R_{app}$:
\begin{equation}
    C_{app}=\dfrac{R_{app}-Z_{in}}{j\omega R_{app}Z_{in}}
\end{equation}
As mentioned in the previous section, the transfer function describing the apparent compliance is frequency-dependent and describes, not only the total arterial compliance but also incorporates other physiological effects such as pulse reflections. At low frequency, $C_{app}$ convergences to a value that approximates  the true total arterial compliance ($C_{tot}$):
\begin{equation}
    C_{tot}=\lim_{w\rightarrow 0}C_{app}
\end{equation}
\subsubsection{Fractional-order capacitor (FOC)}
FOC also known as Constant Phase Element \cite{nakagawa1992basic}, is the main building block for developing analog model structure according to FC. FOC is an electrical element that represents a fractional-order derivative relationship between the current, $i(t)$, passing through and the voltage, $v(t)$, across it with respect to time, $t$, as follow:
\begin{equation}
    i(t)=C_\alpha D_t^\alpha v(t),
\end{equation}
where $C_\alpha$ is a proportionality constant so-called pseudo-capacitance, expressed in units of $\mathrm{[Farad/second^{1-\alpha}]}$. The conventional capacitance, $C$, in unit of $\mathrm{Farad}$ is related to $C_\alpha$ as follow:
\begin{equation}
    C=C_\alpha \omega^{\alpha-1}.
\end{equation}
The impedance ($Z_{FOC}$) of FOC in Laplace domain is given as:
 \begin{equation}
     Z_{FOC}(s)=\dfrac{1}{C_\alpha s^\alpha}. 
 \end{equation}
Substituting the \textit{Laplace} variable, $s$, by ($j\omega$), (9) can be expressed as:
\begin{equation} 
{Z_{FOC}}\left ( \omega \right )=  \underbrace{\dfrac{1}{C_\alpha} \omega^{-\alpha } \cos(\phi)}_\text{\large$G_r$}-j\underbrace{\dfrac{1}{C_\alpha} \omega^{\alpha }\sin(\phi)}_\text{\large $H_r$},
\end{equation}
where $\phi$ represents the phase shift given by the formula: $\phi\!\!=\!\!\alpha{\pi}/ {2}$ [rad] or $\phi\!\!=\!\!90 \alpha$ [degree or $^\circ$]. As illustrated in Fig. S1 in the Supplementary Materials, the bounding values of $\alpha$ represent the discrete conventional elements: the resistor when $\alpha = 0$ and the capacitor when $ \alpha = 1$). Additionally, from (9), it is clear that as $\alpha$ goes to $0$, the imaginary part ($H_r$) of $Z_C$ vanishes to $0$ and hence the FOC characteristic becomes more like that a pure resistor, whereas as $\alpha$ approaches to $1$, the real part ($G_r$) converges to $0$ and hence, FOC operates as a pure capacitors. Furthermore, it has been demonstrated that the characteristics of FOC can be approximated using RC ladder structure \cite{tsirimokou2017systematic}.
Based on the above properties and in comparison to an integer order model where $\alpha$ is strictly fixed to an integer ($0$ or $1$), the parameter $\alpha$ offers extra flexibility for a fractional-order paradigm. In connection with the apparent compliance modeling concept, FOC can be considered as a great candidate that might overcome the discrepancies stemming from integer-order limitation as follows:
\begin{itemize}
    \item The proportionality constant $C_\alpha$ (pseudo-capacitance) is expressed in unit of $\mathrm{[F.sec^{1-\alpha}]}$ that makes, by its very nature, the conventional capacitance $C$, in the unit of $\mathrm{[Farad]}$, frequency-dependent, hence FOC has a physical foundation in representing the complex and frequency dependence of $C_{app}$
    \item Based on the order of the fractional differentiation factor $\alpha$, the storage and the dissipation parts of the resultant FOC's impedance can have different levels,as illustrated in Fig. S2, in the Supplementary Materials. Thus FOC might offer a key advantage in modeling complex system, that is the whole spectrum of dissipative and storage mechanisms may be included in a single parameter (the fractional differentiation order).
    \item The equivalent analog circuit of FOC can be viewed as infinity Voigt cells connected in parallel. Hence FOC might lead to a minimal representation of the mechanical properties of the arterial network by using only two parameters ($\alpha$ and $C_\alpha$).
\end{itemize}
In biorheological research field, the imaginary, as well as the real part of $Z_{FOC}$, might represent the tissue damping ($G_r$) and tissue elastance ($H_r$), respectively: 
\begin{equation}
\left\lbrace
\begin{matrix}
    G_r(j\omega)=\dfrac{1}{C_\alpha \omega^\alpha}cos(\alpha\dfrac{\pi}{2})
\\
    H_r(j\omega)=-\dfrac{1}{C_\alpha \omega^\alpha}sin(\alpha\dfrac{\pi}{2})
\end{matrix}\right.
\end{equation}
The hysteresivity coefficient $\eta_r$ (dimensionless) is defined as:
\begin{equation}
    \eta_r=\dfrac{G_r}{H_r}=-\cot(\alpha\dfrac{\pi}{2})
\end{equation}
In general, these parameters are usually used to characterize the heterogeneity of the bio-tissue and are shown to be variable with pathology, as demonstrated for lung tissue for the respiratory system in \cite{ionescu2013human}.
\subsection{Models}
\subsubsection{Fractional order model of the dynamic Volume/Input-pressure relationship}
Recent researches have shown the key advantages of applying fractional calculus tools to describe correctly: 1) the viscoelasticity properties of the collagenous tissues in the arterial bed, 2) analyze the arterial blood flow \cite{perdikaris2014fractional,zerpa2015modeling} and red blood cell (RBC) membrane mechanics \cite{craiem2010fractional} and, 3) modeling the heart valve cusp \cite{doehring2005fractional}. Bearing this in mind, in this part, we introduce the fractionalization of the dynamic relationship of the arterial blood volume and input-pressure.
Based on the conservation mass, the arterial blood flow pumped from the heart to the arterial vascular bed ($q_{in}$) can be expressed as:
\begin{equation}
    q_{in}(t)=q_{stored}(t)+q_{out}(t),
\end{equation}
where $q_{stored}$ is the blood stored in the arterial tree, and $q_{out}$ corresponds to the flow out of the arterial system. As described in (3), $q_{out}$ can be expressed as:
\begin{equation}
    q_{out}(t)=\frac{1}{R_{app}}p_{in}(t).
\end{equation}
Regarding $q_{stored}$, typically using the conventional definition, it can be determined as the rate of flow by taking the first derivative of the volume equation for the time, whereas, in consideration of the fractional properties of both RBC and the collagenous tissues forming the arterial bed, we allow the differentiation order of the blood volume for time to be real ($\alpha \in [0\ 1]$) and hence applying the fractional-order derivative to this differential equation.     
\begin{equation}
    q_{stored}(t)=D_t^{\alpha} V(t)= \frac{{d^{\alpha}V(t)} }{{dt^{\alpha}}},
\end{equation}
\begin{equation}
    q_{stored}(t)= \underbrace{\frac{d^{\alpha}V(t)}{d^{\alpha}p_{in}(t)}}_{A_{\alpha}} \frac{{d^{\alpha}p_{in}(t)} }{{dt^{\alpha}}},
\end{equation}
where $ A_\alpha$ is a fractional order proportionality constant that can be defined as a fractional order compliance expressed in the unit of $\mathrm{[l/mmHg\ . sec^{1-\alpha}]}$. Assuming null initial condition,the \textit{Laplace} transform of (15) is given as:
\begin{equation}
     Q_{stored}= A_\alpha s^\alpha P_{in}.
\end{equation}
The conventional compliance in the unit of $\mathrm{[l/mmHg]}$ that represents the complex and frequency-dependent apparent compliance can be written as follow:
\begin{equation}
    C_C=A_\alpha s^{\alpha-1}.
\end{equation}
Accordingly, by analogy to the electrical circuit, one may appropriately consider the fractional-order capacitor as a lumped parametric element to stand for the apparent arterial compliance. The voltage is equivalent to the arterial pressure; the electrical charges correspond to the blood volume and the electrical current as of the equivalent of the blood flow. 
\subsubsection{Apparent compliance fractional-order Models}
\begin{figure*}[!t]
\vspace{-.5cm}
	\centering
	\includegraphics[height=5cm,width=17.5cm]{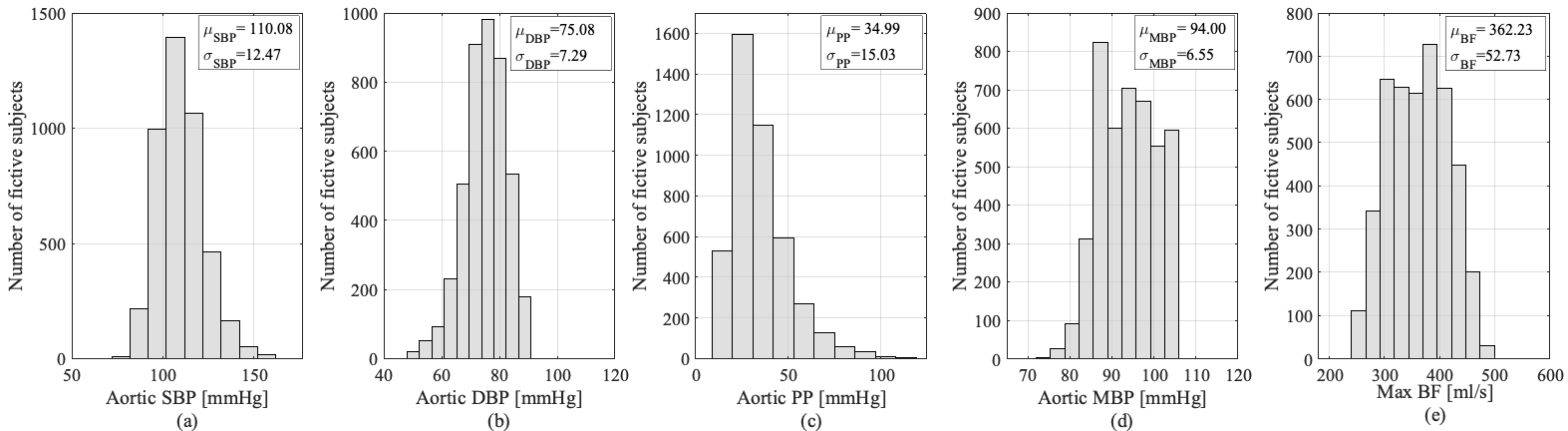}
	\caption{ Distribution, mean value and standard deviation of: (a) systolic blood pressure (SBP), (b) diastolic blood pressure (DBP), (c) aortic pulse pressure ($APP=SBP-DBP$), (d) mean blood pressure (MBP), and (e) the maximum of the blood low (BF) at the level of of ascending aorta for 4,374  virtual subject based in-silico database.}
	\vspace{-.5cm}
\end{figure*}
In this part, we show the derivations of the proposed model based-structures for modeling the apparent arterial compliance. These structures scheme different combinations of FOC along with the conventional resistor and capacitor to display the complex and frequency-dependent behavior of the real dynamic compliance. Fig. 1 shows the proposed electrical analog structures of the proposed models.\\
\textit{\textbf{Model A:}} It comprises only one single FOC. As detailed in the previous sections, the apparent compliance expressed in unit of $\mathrm{[l/mmHg]}$ can be written as:
\begin{equation}
    C_{c}^A=C_\alpha s^{\alpha-1}.
\end{equation}
\textit{\textbf{Model B:}} It comprises a resistor ($R$) and FOC connected in series. The apparent compliance expressed in the unit of $\mathrm{[l/mmHg]}$ can be written as:
\begin{equation}
    C_{c}^B=\dfrac{C_\alpha s^{\alpha-1}}{1+RC_\alpha s}.
\end{equation}
\textit{\textbf{Model C:}} It comprises an ideal capacitor ($  C_{stat}$) accounting for the static compliance and FOC connected in series. The apparent compliance expressed in the unit of $\mathrm{[l/mmHg]}$ can be written as:
\begin{equation}
    C_{c}^C=\frac{C_\alpha C_{stat} s^{\alpha}}{C_\alpha s^\alpha+C_{stat}s}.
\end{equation}
\textit{\textbf{Model D:}} It comprises a resistor ($R$), an ideal capacitor ($C_{stat}$) and FOC connected in series. The apparent compliance expressed in unit of $\mathrm{[l/mmHg]}$ can be written as:
\begin{equation}
    C_{c}^D=\dfrac{ C_{stat}C_\alpha s^{\alpha}}{ C_{stat}s+C_{\alpha}s^\alpha+RC_\alpha C_{stat}s^{\alpha+1}}.
\end{equation}
\textit{\textbf{Model E:}} It comprises a resistor ($R_1$) in parallel to a FOC and a resistor ($R_2$) connected in series. The apparent compliance expressed in unit of $\mathrm{[l/mmHg]}$ can be written as:
\begin{equation}
    C_{c}^E=\frac{ 1+(R_1+R_2)C_\alpha s^{\alpha-1}}{ R_1(1+R_2C_{\alpha}s^\alpha)}.
\end{equation}
\subsection{In-silico Virtual Population}
Owing to a lack of real data to validate the proposed approaches, in this study, we utilize a virtual database of simulated pulse waves (PWs) \cite{charlton2019modelling}. The publicly available PW database \footnote{\url{https://peterhcharlton.github.io/pwdb/index.html}} is considered a useful resource to evaluate the pre-clinical assessment of PWs analysis algorithms. The database encompasses mainly these arterial PWs: 1) flow velocity, 2) luminal area, 3) pressure and 4) photoplethysmogram pulse waves at different sites of the arterial network such as the ascending aorta, carotid artery, brachial artery, and radial arteries. The database represents samples of 4,374 virtual healthy adults aged from 25 to 75 years old, in ten-year increments (six age groups). For each age group, 729 virtual subjects based on pulse waves were created by varying specific cardiac and arterial parameters like the arterial stiffness and heart rate within normal ranges.\\
In this study, PWs at the level of ascending aorta have been investigated to evaluate our approaches. Fig. 2 shows a summary statistic of the aortic blood pressure parameter as well as the maximum blood flow at the level of the ascending aorta, for all virtual subjects. Additionally, we present a detailed statistic summary based on the group age and heart rate in Table SI in the supplementary material. This database presents physiological values with well-balanced distributions.
\subsection{Identification Algorithm}
The parameters of the proposed fractional-order models were estimated by a non-linear least square minimization routine, making use of the well-known $\mathrm{MATLAB-R2019b}$, function \textit{lsqnonlin}. This function is based on the trust-region reflective method \cite{coleman1996interior}. The steps used to obtain the optimal estimates are outlined in Algorithm 1.
\begin{algorithm}[!h]
\caption{Parameter calibration of the apparent compliance models}
\begin{algorithmic}[1]
\State  \text{Load}  the in-silico aortic blood  pressure ($ \mathrm{P}$) and flow  ($ \mathrm{Q}$)
\State  \text{Evaluate} the Fast Fourier Transform (FFT) of both $ \mathrm{ P}$ and $ \mathrm{Q}$
\State  \text{Select} the frequency range (Hz) $ \mathrm{ f\in [0 \ 12]}$ 
\State \text{Calculate} the aortic input impedance $ \mathrm{Z_{in}}$ 
\par  \hskip\algorithmicindent \Comment Using equation (1)
\State \text{Calculate} the in-silico apparent compliance $ \mathrm{ C_{app}}$

\Statex \Comment Using equation (4)
\State Select the model to fit with the data
\State Include and Initialize the parameter to estimate $\mathrm{\Theta}$
\Statex
\% {For instance for a single fractional-order capacitor based model (model A), $ \mathrm{\Theta=\{C_\alpha,\  \alpha\}}$}
\State
\begin{align*}
   \mathrm{ \ RMSE= \sqrt{\frac{\sum_{i=1}^{N_s}\left(\frac{Re-\hat{Re}}{max(Re)}\right)^2+\left(\frac{Im-\hat{Im}}{Im}\right)^2}{N_s}}}
\end{align*}
\Statex
\begin{align*}
   \mathrm{ \hat\Theta=arg \ \underset{\Theta}{min} \ \ RMSE}
\end{align*}
\Statex\% {Where $\mathrm{N_s}$ denoting the number of excited frequency points, $\mathrm{Re}$ and $\mathrm{Im}$ denoting the real and imaginary parts of the real $\mathrm{C_{app}}$, and $\mathrm{Im}$, evaluated in step (5), and $\mathrm{\hat{Re}}$ and $\mathrm{\hat{Im}}$ designate the real and imaginary parts of the model of $\mathrm{C_{app}}$, respectively. $\mathrm{\hat\theta}$ denotes the estimates that minimize $\mathrm{RMSE}$}
\end{algorithmic}
\end{algorithm}
\begin{figure*}[!t]
\vspace{-.5cm}
	\centering
	\includegraphics[height=5.5cm,width=18cm]{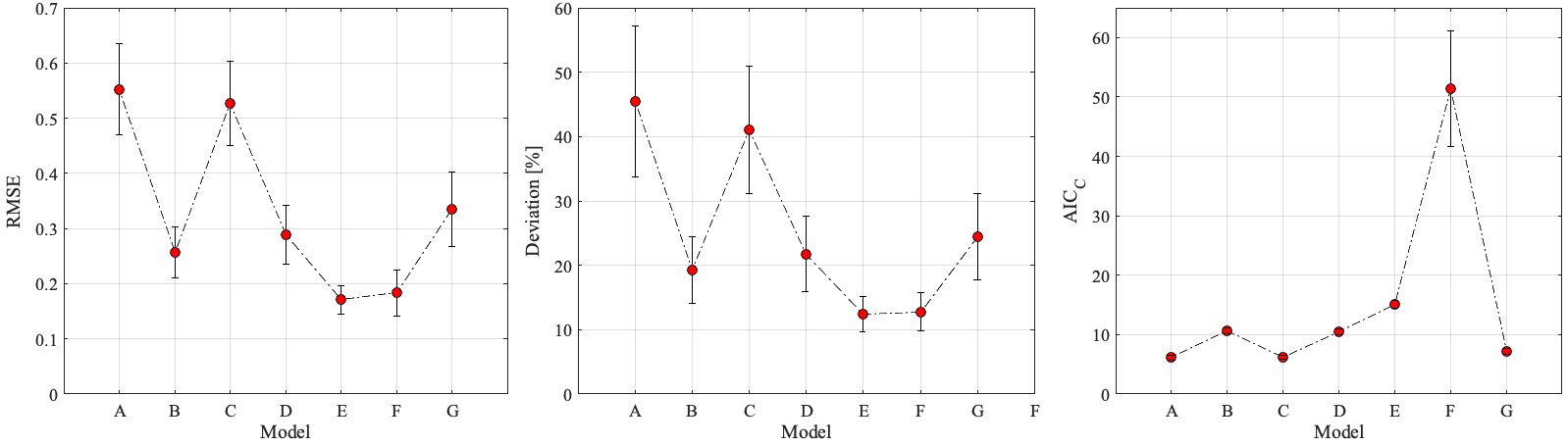}
		\includegraphics[height=4cm,width=18cm]{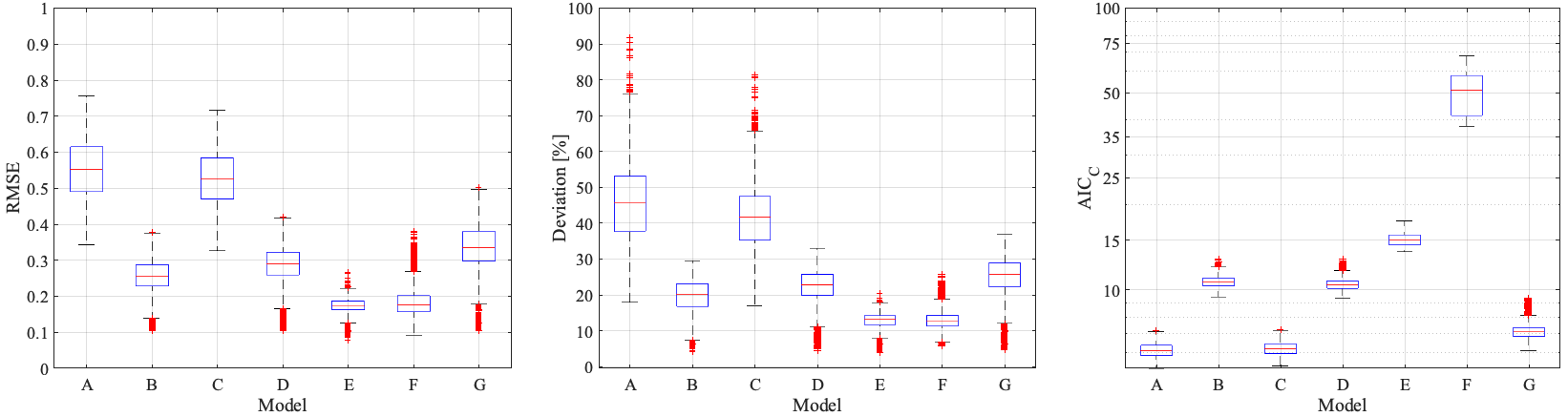}
	\caption{Comparison of goodness of fit quantified as the mean values of RMSE , Devivation, and  , $\mathrm{AIc_c}$ evaluated for both proposed models along with Viscoelastic and Voigt models for all the virtual subjects, and box plots providing a visualization of summary statistics of this comparison.}
	\vspace{-.5cm}
\end{figure*}

In this study, we compare the performance of the proposed model with their corresponding integer-order version, where the fractional differentiation order $\alpha$ is equal to $1$. Accordingly, the integer order version of model \textbf{A} will be equivalent to an ideal capacitor whose capacitance is a constant, which is not frequency-dependent, hence in our comparison, we exclude this case. Similarly, the integer-order version of model \textbf{C} leads to a series of two ideal capacitors, which is as well, equivalent to an ideal capacitor whose capacitance is constant.
For the models, \textbf{B} and \textbf{D}, the integer-order version is equivalent to the analog Voigt cell model (an ideal capacitor connected in series with a resistor). Conclusively, in this work, we compare our proposed methods to the Voigt cell model. 
On top of that, in order to show the role of fractional-order concept in reducing the complexity of such approach, we conduct an extra comparison with the well-known general apparent compliance-based model for viscoelastic material \cite{goedhard1973model} which is expressed as:
\begin{equation}
    C_{c}^F=C_{stat} \dfrac{\prod_{n=1}^{N}a_n(j\omega+b_n)}{\prod_{n=1}^{N}b_n(j\omega+a_n)},
\end{equation}
where $a_n$ and $b_n$ are imperial constants that can be convenient to fit any particular case. $C_{stat}$ denoting the static compliance for the vessel. Goedhard et. all showed that this model could fit an experimental data with $N\!\!=\!\!4$. Hence in our comparison we choose $N\!\!=\!\!4$. We refer to the viscoelastic model and Voigt model as models \textbf{F} and \textbf{G}, respectively.
Because the proposed models have a different number of parameter, to perform a fair comparison, the corrected Akaike Information Criterion ($AIC_C$) was evaluated:
\begin{equation}
    AIC_C=-2ln(RMSE)+\frac{2PN_s}{N_s-P-1},
\end{equation}
where P is the number of parameters. Furthermore, the deviation of the model modulus from the in-silico apparent compliance modulus was calculated, using the following expression:
\begin{equation}
D_i\ [\%]=\left [ \dfrac{\left | {C}_{c_{[i]}}^{model} \right |-\left| {C_{app_{[i]}}} \right |}{\left| C_{app_{[i]}} \right |}\right]_{i=1..N_s} \times 100 \%.
\end{equation}
For ease of visualization of the various comparisons between the different models, for each virtual subject, we evaluated the mean of D [\%] over the $N_s$ harmonics, based on the following equation:
\begin{equation}
Deviation\ [\%]=\dfrac{\sum_{i=1}^{N_s}D_{i} [\%]} {N_s}.
\end{equation}
\section{Results and discussion}
In this section, we first present a comparative evaluation between the proposed fractional-order model along with the integer-order ones. Afterward, we discuss the parameter estimates and their physiological insights. 
\subsection{Quantifying the models performances}
The mean values of the goodness of fit criterion
($\mathrm{NRMSE}$, $\mathrm{{Deviation}(\%)}$ and $\mathrm{{AIC_c}}$), after applying all the models, are depicted in Fig. 3, along with their box plots providing a visualization of summary statistics of this comparison. Additionally, we listed in Table SII, in the supplementary material, all the $\mathrm{RMSE}$, $\mathrm{Deviation}$ and  $\mathrm{AIC_C}$ mean values for each group of age and heart rate of the in-silico data. Fig. S3, in the Supplementary Materials represents a map of all the models with respect to the $\mathrm{Deviation}$ and the number of parameters to estimate (complexity).
As for any modeling system, based on the fitting performance results, it is clear that there is a good compromise between the accuracy and complexity of the model. Among the five proposed fractional-order model, the single fractional-order capacitor-based, \textbf{Model A}, failed to produce the lowest $\mathrm{RMSE}$ and $\mathrm{Deviation}$ for any of the data sets; however, it represents the smallest ($\mathrm{AIC_C}$). The fractional-order \textbf{Model E} and the integer-order \textbf{Model F} exhibit the lowest $\mathrm{Deviation}$, as well as $\mathrm{RMSE}$ at the expense of the complexity that was reflected in the highest values of $\mathrm{AIC_C}$. It is worth to mention that the fractional-order \textbf{Model E} comprises only four parameters and performed better than \textbf{Model F}, which posses nine parameters. As illustrated in Fig.7, the best models that compromise between the complexity and the accuracy are \textbf{Model B}, \textbf{Model D}, and \textbf{Model G}. In terms of accuracy performance, among the latest models, the fractional-order \textbf{Model B} is performing the optimal.
Conclusively, from the previous analysis, it is apparent that model system fractionalizing is enhancing the accuracy of the arterial compliance as well as reducing the complexity.
\subsection{Statistical Analysis of the estimated parameters}
\subsubsection{Model A}
\begin{figure}[!b]
	\vspace{-.5cm}
	\centering
	\includegraphics[height=5cm,width=8cm]{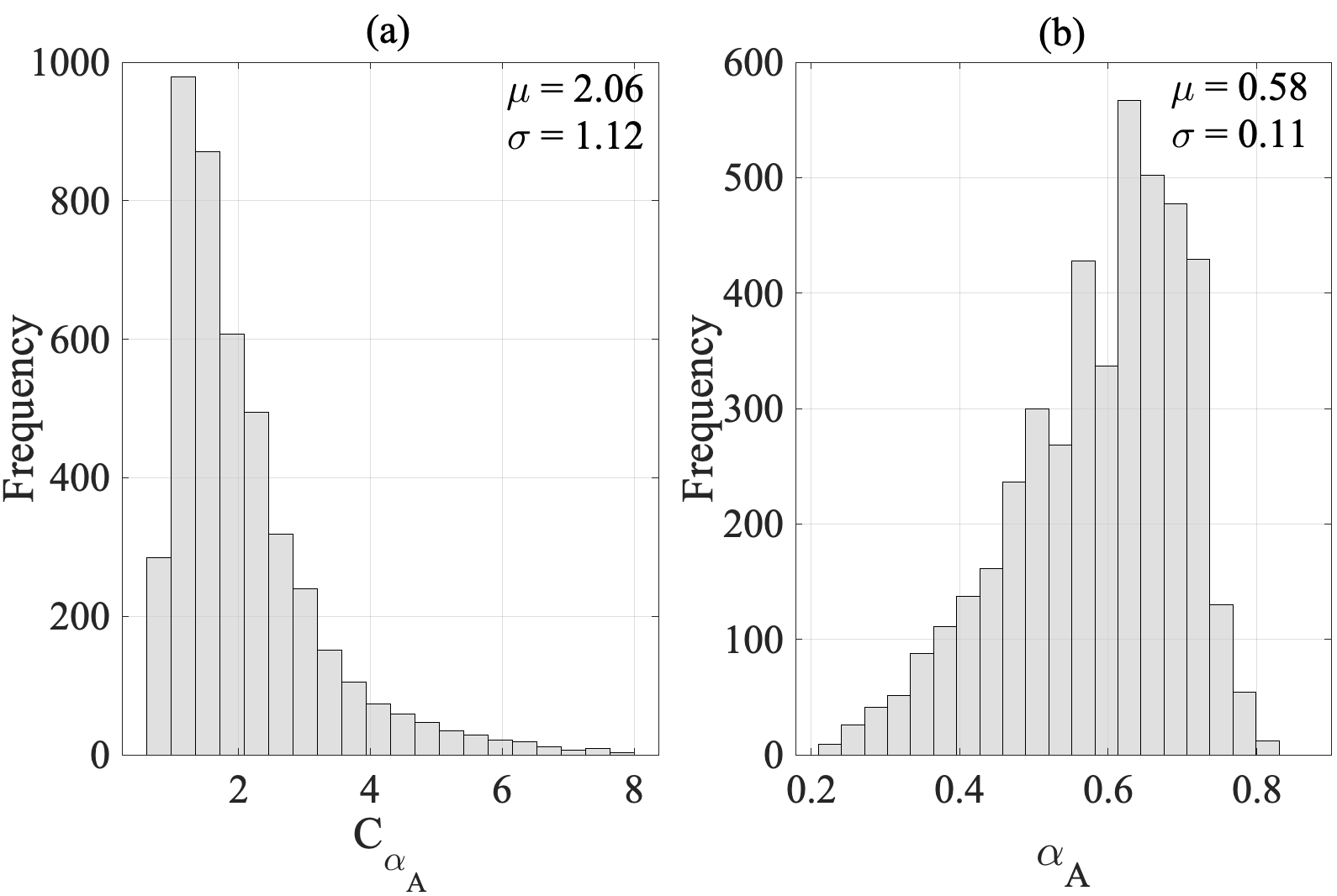}
	\caption{Distribution, mean value and standard deviation of Model A parameter estimates: (a) the pseudo-capacitance $C_{\alpha_A}$, and (b) the fractional differentiation order parameter $\alpha_A$.}
\end{figure}
\begin{figure*}[!t]
	\vspace{-.5cm}
	\centering
	\includegraphics[height=5cm,width=13cm]{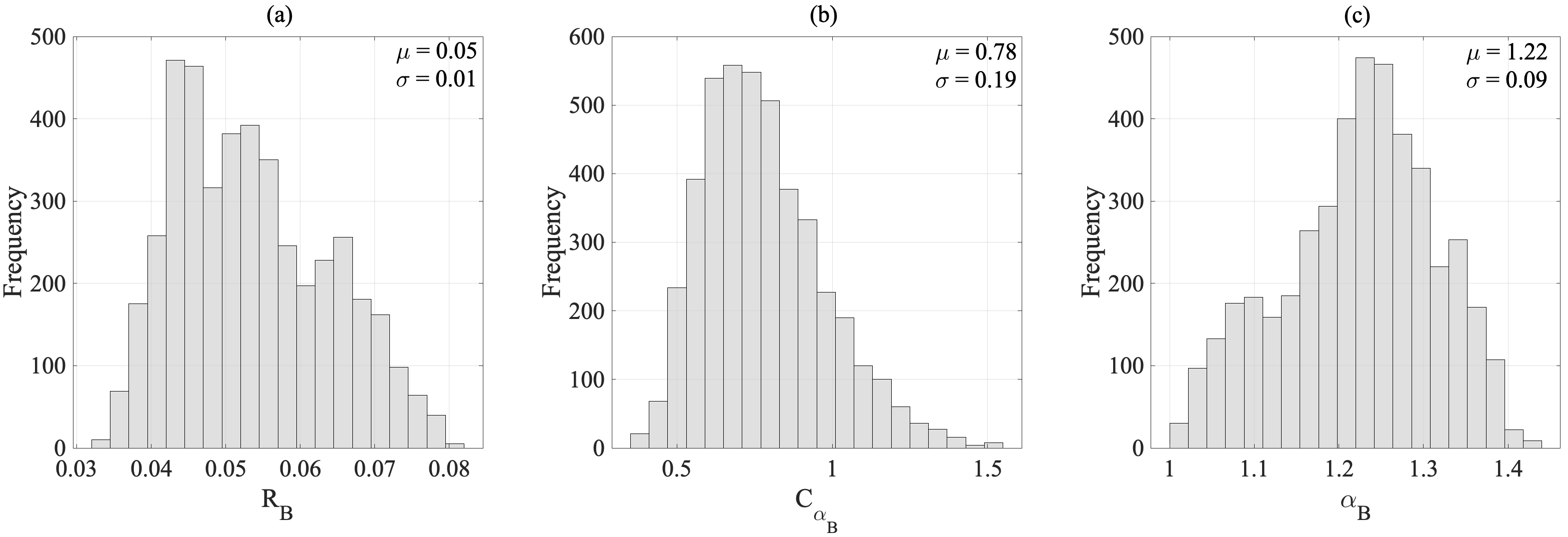}
	\caption{Distribution, mean value and standard deviation of Model B parameter estimates: (a) the resistance $R_B$, (b) the pseudo-capacitance $C_{\alpha_B}$, and (c) the fractional differentiation order parameter $\alpha_B$.}
		\vspace{-.5cm}
\end{figure*}
\begin{figure}[!b]
	\vspace{-.5cm}
	\centering
	\includegraphics[height=5cm,width=8cm]{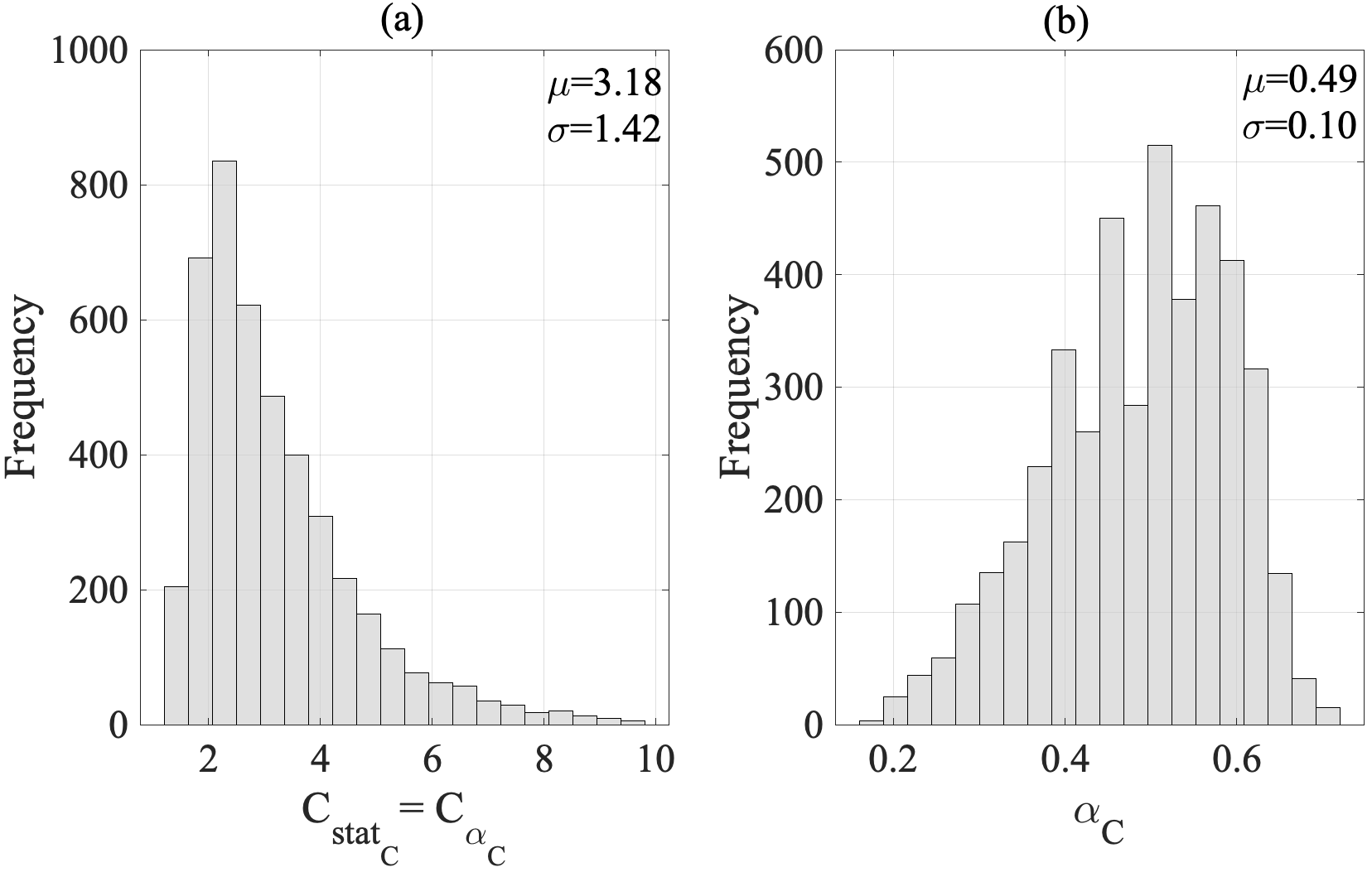}
	\caption{Distribution, mean value and standard deviation of Model C parameter estimates: (a) the static capacitance $C_{stat}$ and the pseudo capacitance $C_{\alpha_C}$ which are equals, and (b) the fractional differentiation order parameter $\alpha_C$.}
\end{figure}
\begin{figure*}[!t]
	\vspace{-.5cm}
	\centering
	\includegraphics[height=5cm,width=14cm]{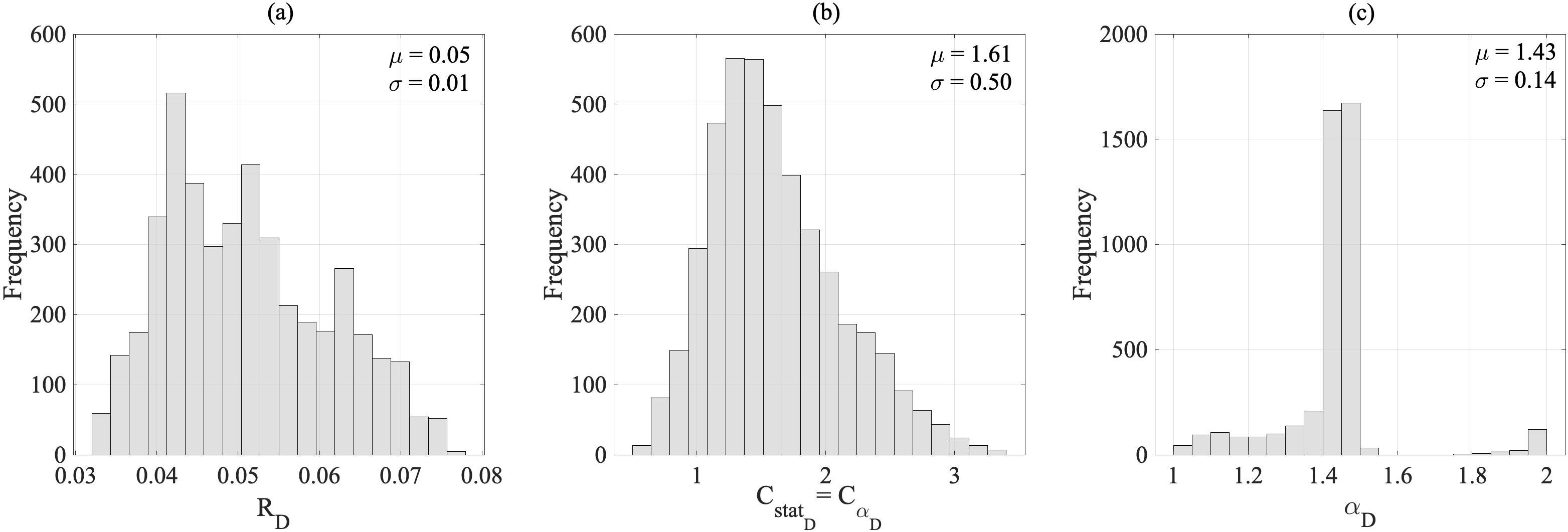}
	\caption{Distribution, mean value and standard deviation of Model D parameter estimates: (a) the resistance $R_D$, (b) the static capacitance $C_{stat_D}$ and the pseudo capacitance $C_{\alpha_D}$ which are equals, and (c) the fractional differentiation order parameter $\alpha_D$.}
		\vspace{-.5cm}
\end{figure*}
\begin{figure*}[!b]
	\centering
	\includegraphics[height=5cm,width=18cm]{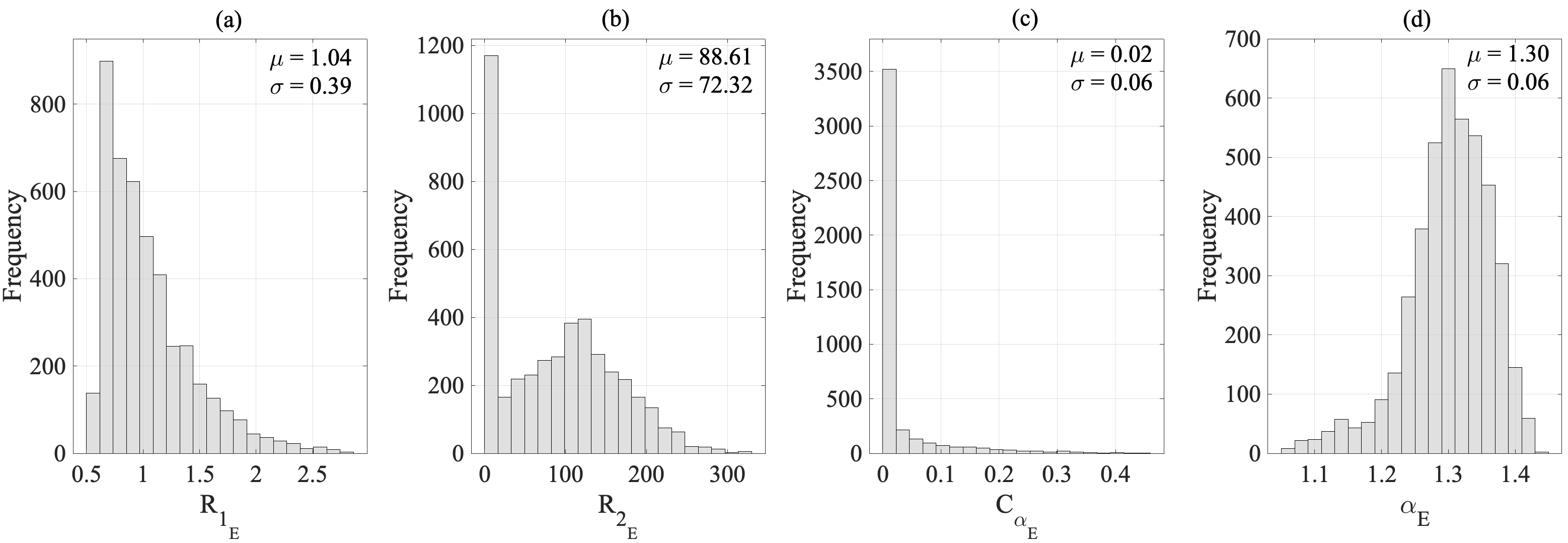}
	\caption{Distribution, mean value and standard deviation of Model E parameter estimates: (a) the resistance $R_{1_E}$, (b) the resistance $R_{2_E}$, (c) the pseudo capacitance $C_{\alpha_E}$, and (d) the fractional differentiation order parameter $\alpha_E$.}
\end{figure*}
Fig. 4 shows the distribution of the parameter estimates of Model A, after fitting the in-silico data of the arterial compliance. By observing the distribution of the fractional differentiation order, $\alpha$, estimates, it is clear that this parameter is less than $1$ for all the subjects. Its mean value is approximately $0.58 \pm 0.008$. It is worth noting that in the estimation phase, for the parameter $\alpha$, we have only constrained the lower bound to be zero; however, for the upper bound, it was unconstrained. Accordingly, this result indicates that the arterial system exhibits a viscoelastic behavior, not a purely elastic one. Indeed, the fact that $\alpha \neq 1 $ implies that the FoC element incorporates both resistance and capacitance behaviors, as demonstrated mathematically in (10). This result further supports the concept of fractional-order behavior by the arterial system. In the proposed model, the fractional-order element combines both the resistance and the capacitance properties,  which display the viscoelastic behavior of the arterial vessel. The contributions from both properties are controlled by the fractional differentiation order $\alpha$, enabling a more flexible physiological description.  As the fractional power approaches to $1$, the capacitance part dominates and, hence the arterial system behaves like a pure elastic system.
\subsubsection{Model B}
Fig. 5 (a), (b), and (c) show the distribution of the parameter estimates of the Model B after fitting the in-silico data of the arterial compliance. By observing the distribution of $\alpha_B$, it is clear that for all the subjects, this parameter is higher than $1$  with a mean value of approximately equals to $1.22\pm0.09$. Mathematically, as $\alpha$ exceeds $1$, the real part of the fractional-order element impedance, $\mathrm{G_r}$, becomes negative, and hence it has the characteristic of a negative resistor that supplies power. Having a negative resistance, in this case, comes as compensation for the added series resistance $\mathrm{R_B}$. Besides, comparing to Model A, it is worthy to notice that the mean value of the pseudo-capacitance $\mathrm{C_{\alpha_B}}$ was decreased to $\mathrm{0.78 \pm 0.19}$.          
\subsubsection{Model C}
Fig. 6 (a) and (b) show the distribution of the parameter estimates of the Model C after fitting the in-silico data of the arterial compliance. In this model, the static capacitance has been chosen to be equal to the pseudo-capacitance. By observing the distribution of $\mathrm{\alpha_C}$, we can notice that this parameter is less than $\mathrm{1}$ for all the population with a mean value equal to $\mathrm{0.49 \pm 0.10}$. Comparing to Model A, the fractional factor has been decreased by approximately $\mathrm{0.1}$; however, the pseudo-capacitance was increased by $\mathrm{1}$. The decrease of $\alpha$ implies an increase of the resistive part of the FOC, as explained in the previous parts, which comes as to compensate for the increase in the overall capacitive part of the whole system model. In this model, the ideal capacitor is counting for static compliance, whereas the fractional-order one controls the arterial stiffness level. In other words, $\alpha$ might give a piece of information about the variation of the viscoelasticity of the arteries. 
\subsubsection{Model D}
Fig. 7 (a) and (b) show the distribution of the parameter estimates of the Model D after fitting the in-silico data of the arterial compliance. This model incorporates a static capacitor in series to FOC, along with a small resistor. Comparing to Model C, the addition of the small resistance causes $\alpha$ to go beyond $1$ with a mean value approximately equal to $\mathrm{1.43 \pm 0.14}$. Besides, the mean value of static compliance and pseudo-capacitance decreases to $\mathrm{1.61\pm0.50}$. The addition of serial constant resistor and capacitor in this model is for the sake of account for the static viscosity and elasticity, respectively, while FOC depicts the ability of the arterial vessel to store blood dynamically.
\subsubsection{Model E}
Fig. 8 (a), (b), (c), and (d) show the distribution of the parameter estimates of the Model E after fitting the in-silico data of the arterial compliance. This model is similar to the equivalent analog circuit of Maxwell's mechanical element ( series spring and dashpot in parallel with a dashpot), whereas instead of using an ideal capacitor to represent the spring, FOC has been employed. In terms of performance, Model E is the best. Similar to all the proposed model, $\alpha_{D} \in \mathbb{R}$ which demonstrate the fractional-order behavior of the apparent arterial compliance. By observing the distribution of $\mathrm{R_{1_E}}$ $\mathrm{R_{2_E}}$, it is noticeable that these parameters are larger than $\mathrm{R_B}$ and $\mathrm{R_D}$.
\subsection{Relations between fractional-order parameters and central hemodynamic characteristics}
Several research studies have observed that the changes in the determinants of the central blood pressure waveform, such as systolic blood pressure $\mathrm{SBP}$, diastolic blood pressure $\mathrm{DBP}$, and pulse pressure (APP), are strongly associated with cardiovascular diseases incidents. For instance, the augmentation of the $\mathrm{SBP}$ or $\mathrm{APP}$ is considered as a reflection marking the improper functioning of the cardiovascular system. In fact, stiffer arteries resulting from the arteriosclerosis disease causes increases in the $\mathrm{SBP}$ as well as arterial pulse wave velocity $\mathrm(PWV)$. $\mathrm{PWV}$, such as carotid-to‐femoral one $\mathrm(PWV_{cf})$, are recognized as valuable surrogates of the arterial stiffness. In this part, we investigate whether the fractional differentiation order, $\alpha$, and the hysteresivity coefficient, $\eta_r$, (defined by (11)) correlate with the central blood pressure determinants and the arterial pulse wave velocity $\mathrm(PWV_{a})$ and $\mathrm(PWV_{cf})$. In addition to the arterial pulse waves, the used database provides both $\mathrm(PWV_{a})$ and $\mathrm(PWV_{cf})$ for each subject. 

Accordingly, to evaluate the associations between the fractional-order parameters and the central hemodynamic determinants, for each model, we calculated the average value of $\alpha$ and $\eta_r$ estimates over a fixed interval of the blood pressure determinant $\mathrm{(SBP,\ DBP,\  and\ APP)}$ that is equal to $\mathrm{5\ [mmHg]}$ and $\mathrm{PWV}$ that is equal to $\mathrm{0.5 \ [m/s]}$. Table I shows the correlation coefficients between $\alpha$, and $\eta_r$ and (SBP, DBP, APP, $\mathrm{PWV_{a}}$ and $\mathrm{PWV_{cf}}$) ($95\%$ confidence interval). It is clear from these results that, for the majority of the models, the fractional-order parameters are strongly associated with the hemodynamic determinants. For instance, with regards to $\mathrm{SBP}$, $\mathrm{DBP}$,and $\mathrm{APP}$, we notice that for all the proposed model, excepting for \textbf{Model D} the correlation coefficients with respect to $\eta_r$ and $\alpha$ are larger or equal to $0.90$. Regarding to the stiffness indexes, $\mathrm(PWV_{a})$ and $\mathrm(PWV_{cf})$, the coefficients correlation are approximately equal or larger than $0.85$ for all the proposed model apart from \textbf{Model D}. Overall, the fractional-order parameter estimates of \textbf{Model B} present the best correlation coefficients. This result is in agreement with the goodness of the fit performance of this model. In fact, as analyzed in the previous parts, \textbf{Model B} provides a compromise between the accuracy and complexity of other proposed fractional-order models. In addition, although \textbf{Model A} is not very accurate in estimating the apparent compliance and not complicated, the correlation coefficients between its parameter estimates and the hemodynamic determinant as well as the central $\mathrm{PWVs}$ are acceptable and reasonable. 

Conclusively, our findings point out the potential interests of using FOC in the characterizing of arterial compliance. In addition, it demonstrates the viability of the fractional-order differentiation order to serve as a surrogate measure of the arterial stiffness or marker of cardiovascular diseases. Indeed, by assessing the fractional factor, $\alpha$, it is easy to evaluate the hysteresivity coefficient $\eta_r$ reflecting the ratio between two physiologically insightful parts: the tissue resistance and elastance. 
\begin{table*}[!t]
\centering
\caption{Correlation coefficients between $\alpha$ and arterial systolic blood pressure (SBP),arterial diastolic blood pressure (DBP), arterial pulse pressure (APP), arterial pulse wave velocity ($PWV_{a}$) and carotid-femoral pulse wave velocity ($PWV_{cf}$) ($95\%$ confidence interval)}. 
\begin{tabular}{|c|c|c|c|c|c|}
\hline
           & SBP                        & DBP                       & APP                       & $\mathrm{PWV_a}$                    & $\mathrm{PWV_{cf}}$                     \\ \hline\hline
$\alpha_A$ & 0.96  ( 0.88 , 0.99)       & -0.94   (-0.98  ,  -0.82) & 0.92 (0.78  ,  0.97)      & 0.85  (0.61  ,  0.95)     & 0.87  (0.65  ,   0.95)      \\ \hline\hline
$\eta_{r_A}$ & 0.95   ( 0.85  ,  0.98) & -0.94  ( -0.98 ,  -0.83)   & 0.90    (0.74  ,  0.97)  & 0.84   (0.58  ,  0.94)   & 0.85   (0.62  ,  0.95)   \\ \hline\hline
$\alpha_B$ & -0.99  (-1.00  ,  -0.96)  & 0.96 ( 0.88  ,  0.99)     & -0.97 (-0.99  ,  -0.91)   & -0.98  (-0.99  ,  -0.93)  & -0.97   (-0.99  ,  -0.92)   \\ \hline\hline
$\eta_{r_B}$ & -0.99 (-1.00 ,  -0.96)    & 0.96    (0.89  ,  0.99)    & -0.96   (-0.99 ,  -0.89) & -0.97   (-0.99 ,  -0.92) & -0.97   (-0.99  , -0.90) \\ \hline\hline
$\alpha_C$ & 0.97 (0.91  ,  0.99)       & -0.93  ( -0.98  ,  -0.81) & 0.94  (0.82  ,  0.98)     & 0.88  ( 0.68  ,  0.96)    & 0.89    (0.72  ,  0.96)     \\ \hline\hline
$\eta_{r_C}$ & 0.96   ( 0.87  ,  0.99)   & -0.93   (-0.98  , -0.81) & 0.92  (  0.77  ,   0.97) & 0.86   ( 0.64  ,  0.95)  & 0.87    (0.66  ,  0.96)  \\ \hline\hline
$\alpha_D$ & -0.92  (-0.97  ,  -0.78)   & 0.79    (0.48  ,  0.92)   & -0.89   (-0.96  ,  -0.71) & -0.79 (-0.92  ,  -0.48)   & -0.68   (-0.88  ,  -0.28)   \\ \hline\hline
$\eta_{r_D}$ & 0.35   (-0.20 ,   0.73)   & -0.65   (-0.87  , -0.22)   & 0.34  ( -0.18  ,  0.72)  & 0.25   (-0.28  ,  0.66)  & 0.42  ( -0.09  , 0.76)   \\ \hline\hline
$\alpha_E$ & -0.96  ( -0.99   ,  -0.88) & 0.95  (0.85  ,  0.98)     & -0.99  (-1.00  ,  -0.97)  & -0.90   (-0.97  ,  -0.74) & -0.89  ( -0.96  ,  -0.71) \\ \hline\hline
$\eta_{r_E}$ & -0.96  (-0.99 ,  -0.89)   & 0.95    (0.86  ,  0.98)    & -0.99   (-1.00   -0.97)  & -0.91   (-0.97  , -0.75) & -0.89   (-0.96  , -0.70) \\ \hline
\end{tabular}
\vspace{-.5cm}
\end{table*}
\subsection{Limitations}
The  fractional-order paradigm proposed  in this work  should  be developed  a  little  further  before  its generalization  in  the hemodynamic  modeling  context. In fact, It is worthy to note the limitations of this study. Firstly, in this work, due to the non-availability of real data, we used \textit{in-silico} data. Although this database mimics the real physiological human states, and it is based on a validated one-dimensional numerical model of the arterial network, \textit{in-vivo} investigations are required to validate and verify the reliability of the proposed models. The  use  of  real data would considerably give more credibility to the new paradigm.  
Secondly, the estimation was based on only one cardiac cycle. Future work should derive metrics from multiple cycles. This will help to assess and take into account the inter-beat interval variability.

In addition, the  presented  approaches  should  be  conducted  in a  range  of  different  real  physiological  situations  and  show  a good fitting for all the cases. It  is  straightforward  to use  FOC  in  the  simple  model  representation  proposed  here, but  there  is  no  explicit  agreement  on the  exact  physiological relevance  of  the  new  parameter,  the  fractional  differentiation order $\alpha$ or $\eta_r$.  Although  it  is  evident  from  mathematical  equations that $\alpha$ value  controls  the  viscosity  as  well  as  the  elasticity levels,  it  would  be  of  great  potential  for clinical  application, to  define  ranges  of  the  $\alpha$ value  for  normal  and  pathological  physiological  conditions.
Finally, this study does not consider the noise effect on the pulse wave signals. In fact, several sources of noise can be allocated with the blood pressure signal, such as the movement artifacts, poor sensor contact, and optical interference, etc. Accordingly, considering the noise can impact the utility of the estimation of the fractional-order parameters. In the future, the robustness of the parameter estimates against the different sources of noise should be studied and analyzed. This is extremely important to especially assess arterial stiffness. 
\section{Conclusion}
The appearance of fractional-order behavior in the arterial system has been identified by many experimental studies of the viscoelasticity properties of the collagenous tissues in the arterial bed; the analyzes of the arterial blood flow and red blood cell membrane mechanics and the characterizing the heart valve cusp. This paper introduced a fractional-order modeling approach to assess the apparent arterial compliance. The models incorporate FOC along with ideal resistors and capacitors to display the dynamic relationship between the blood volume and aortic input pressure. The majority of proposed parametric models present reasonable fit performance with in-silico data. The results show that fractional-order model structures conveniently capture the capacity of the arterial system to store the blood. Besides, the fractional-order parameter estimates present good correlation coefficients with the central hemodynamic determinants, such as the systolic and pulse blood pressure along with the central pulse wave velocity indexes. Thus, the fractional-order based approach of arterial compliance has a great potential to provide a new alternative in assessing the arterial stiffness. Future investigations will be directed toward integrating these models within a complete lumped-parameter model for the systemic circulation and study the effects of certain cardiovascular pathologies upon changes in the dynamic arterial compliance represented by the fractional-order capacitor.
\section*{Acknowledgment}
Research reported in this publication was supported by King Abdullah University of Science and Technology (KAUST) Base Research Fund (BAS/1/162701-01). Additionally, the authors would like to thank Dr. Ali Haneef, associate consultant cardiac surgeon and co-chairman quality management at King Faisal Cardiac Center, King Abdulaziz Medical City, National Guard Health Aﬀairs, in the Western Region, Jeddah, KSA and Dr. Nesrine T. Bahloul, medical intern at Department of Pediatrics, Sfax Medical School, Hedi Chaker Hospital, Sfax, Tunisia, for their assistance and valuable advices.
\section*{Supplementary Materials}
The Supplementary Information is described in more detail in the Supplementary Materials file, which includes the following Sections: (A) Fractional-order calculus section provides an overview about the Fractional-order derivative and its mathematical definitions; (B) Figures section represents Fig. S1: The schematic diagram for the resistor, capacitor, and fractional-order capacitor elements along with theirs \textit{i-v} characteristic relationships; Fig. S2: Modulus of (FOC impedance $Z_C$, left side), (the dissipation part $Z_D$) and (the storage part $Z_S$) for $C_\alpha=1$; and Fig. S3: A map of all the models with respect to the Deviation and the number of parameters to estimate (complexity), and (C) Tables section presents three tables: Table SI: Mean value of systolic blood pressure (SBP), diastolic blood pressure (DBP), aortic pulse pressure ($\mathrm{APP=SBP-DBP}$), mean blood pressure (MBP), and the maximum of the blood low (BF) at the level of of ascending aorta for 4,374  virtual subject based in-silico database; Table SII: Mean values of the goodness of fit criterion
($\mathrm {NRMSE}$, $\mathrm {Deviation [\%]}$, and $\mathrm {AIC_c}$ of each age and heart rate based-group, and Table SIII: Mean value of the parameter estimates of the fractional-order models for each age and heart rate based-group.
\bibliographystyle{IEEEtran}
\bibliography{bibfile}
\newpage

\title{Fractional-order Modeling of the Arterial Compliance: An Alternative Surrogate Measure of the Arterial Stiffness
}
\author{Mohamed~A. Bahloul and Taous-Meriem Laleg Kirati}
\maketitle
\section*{\textbf{Supplementary Materials}}
\subsection{Fractional calculus}
Over recent decades, the theory of fractional calculus has gained a significant research interest in the field of biology \cite{magin2006fractional}. This is originated from the interdisciplinary nature of this field as well as the flexibility and effectiveness of FC in describing complicated physical systems. For example, the characterization of bio-impedance, modeling of the viscoelasticity and biological cells, and representing the mechanical properties of the arterial system, as well as respiratory systems, have been investigated extensively through the exploring of FC. The concept of FC is not new dating from the pioneer conversation between \textit{L'Hopital} and \textit{Leibniz} in $1695$ that yielded to the generalization of the conventional integer derivative to a non-integer order operator \cite{podlubny1998fractional}, as follow:
 \begin{equation}
 D^{\alpha}_{t}=
 \left\{ \begin{matrix}\dfrac{\mathrm{d^{\alpha}} }{\mathrm{d} t^{\alpha}} 
 & \mbox{if} & \alpha>0 \\ 
 1,& \mbox{if} & \alpha=0,\\ 
 \int_{t}^{0}\left ( df \right )^{-\alpha}& \mbox{if} & \alpha<0
 \end{matrix}\right.
 \end{equation}
where ${\alpha \in \mathbb{R}}$ is the order of the operator known as the fractional-order, and ${df}$ is the derivative function. Numerous fractional calculus definitions have been suggested. Generally, these definitions can be classified into two main classes. In the first class, the operator $  D_t^\alpha$ is converted into the standard differential-integral operator when $\alpha$ is integer. For instance, according to the $Reimann$-$Liouville$ definition, the fractional-order derivative $\alpha$ of a function $g(t)$ can be formulated \cite{jumarie2006modified} as follow:
 \begin{equation}
 D_t^{\alpha}g\left ( t \right )=\frac{1}{\Gamma \left ( 1-\alpha  \right )}\frac{\mathrm{d} }{\mathrm{d} t}\int_{0}^{t}\frac{g\left ( \tau  \right )}{\left ( 1-\tau  \right )^{\alpha }}d\tau, 
 \end{equation}
 where $\Gamma$ is the Euler gamma function. The second class is that the Laplace transform of $ D_t^\alpha$  is $ s^\alpha$, assuming a null initial fractional conditions.The fractional operator is given by:
 \begin{equation}
 D_t^{\alpha}g\left ( t \right )\overset{L}{\rightarrow}s^{\alpha}G\left ( s \right ),
 \end{equation}
 This class is very interesting in developing parametric models for complex systems and control design in frequency domain.
 where $s$ is the complex Laplace. The Fourier transform can be found by substituting $s$ by $jw$ and thus the equivalent frequency-domain expression of  $s^n$ are: 
 \begin{equation}
\left ( j \omega \right )^\alpha=\omega^\alpha\left ( cos\frac{\alpha\pi}{2} -jsin\frac{\alpha\pi}{2}\right ),
\end{equation}
 \begin{equation}
 \frac{1}{\left ( j\omega \right )^\alpha}=\frac{1}{\omega^\alpha}\left ( cos\frac{\alpha\pi}{2} +jsin\frac{\alpha\pi}{2}\right ).
\end{equation}
 \subsection{Figure}
\begin{figure*}[!h]
	\centering
	\includegraphics[height=5cm,width=10cm]{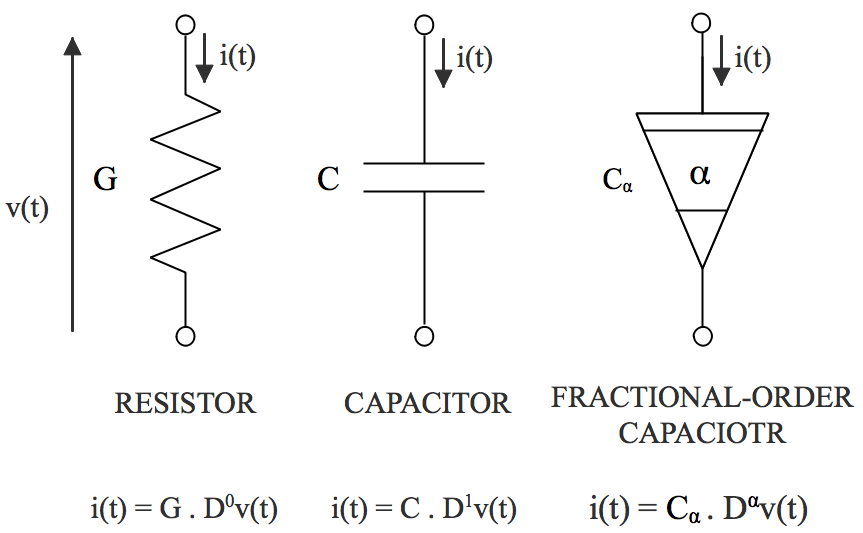}%
	\caption{The schematic diagram for the resistor, capacitor, and fractional-order capacitor elements along with theirs \textit{i-v} characteristic relationships. 
	Here, for the fractional-order capacitor, $ {i(t)=C_{\alpha}D^\alpha v(t)}$ where ${0\leq \alpha \leq1}$ and $C_{\alpha}$ is the pseudo-capacitance. The bounding values of $\alpha$ represent the discrete conventional elements (Resistor, ${C_\alpha = G} $ when $\alpha = 0$; Capacitor, ${C_\alpha = C} $, the capacitance, when $ \alpha = 1$).}
	\label{fig1}
\end{figure*}
 \begin{figure*}[!h]
	\centering
	\includegraphics[height=7cm,width=18cm]{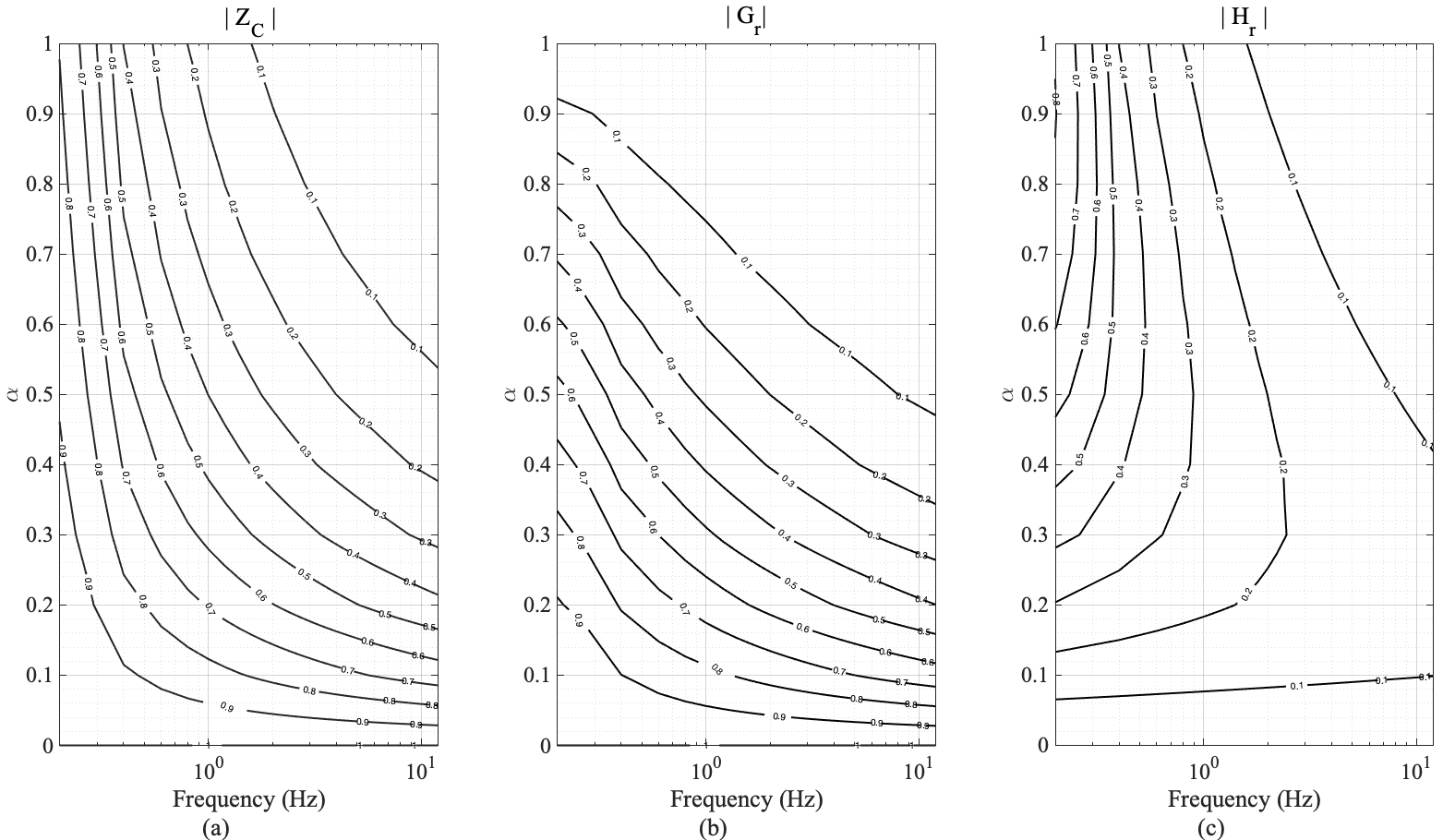}
	\caption{ Modulus of (FOC impedance $Z_C$, left side), (the dissipation part $Z_D$) and (the storage part $Z_S$) for $C_\alpha=1$.}
	\label{fig2}
\end{figure*}
\begin{figure*}[!h]
	\centering
	\includegraphics[height=8cm,width=12cm]{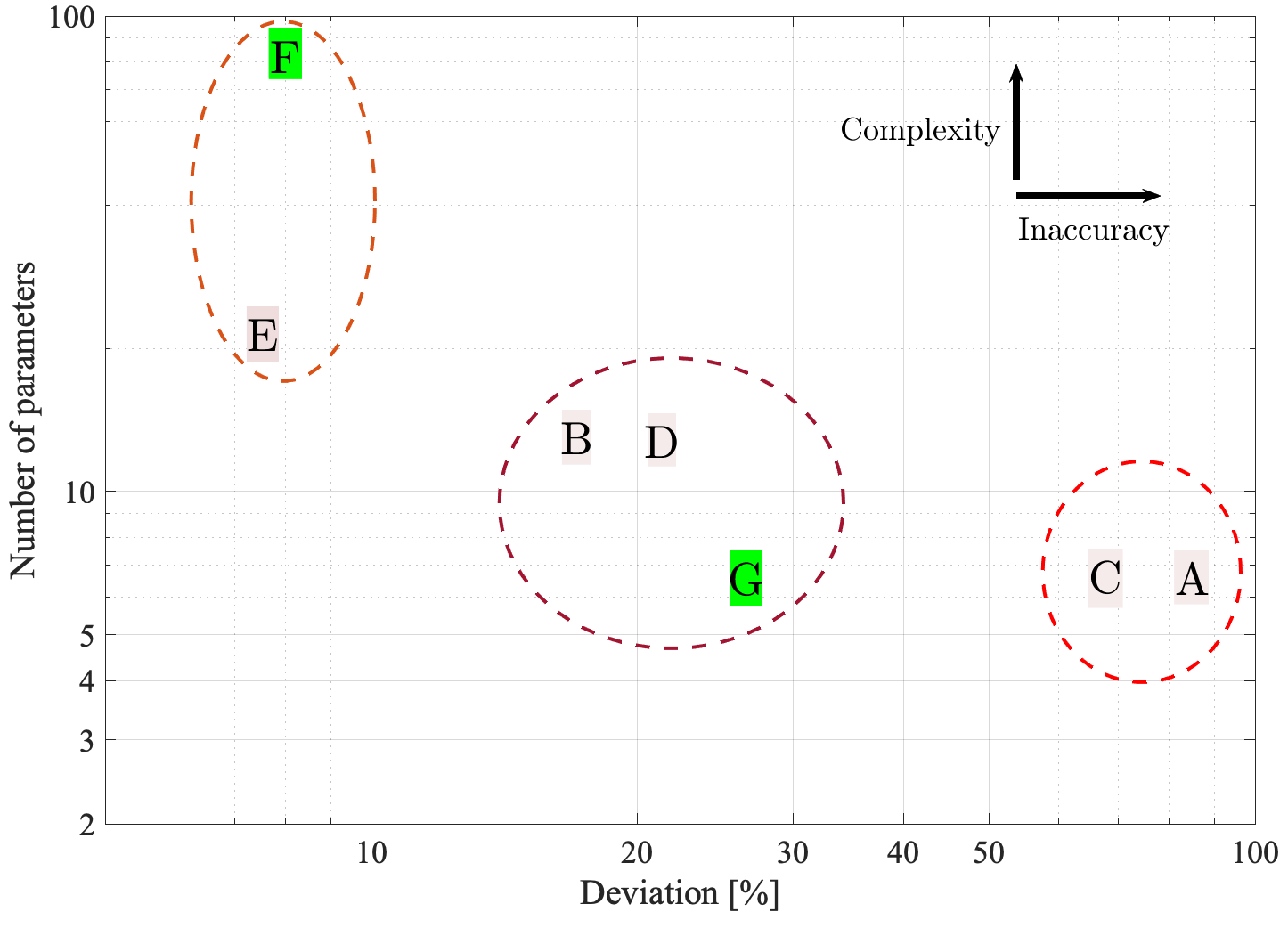}
	\caption{A map of all the models with respect to the $\mathrm{Deviation}$ and the number of parameters to estimate (complexity).}
\end{figure*}
\subsection{Table}
\begin{landscape}
\begin{table}[p]
\ra{1.5}
	\centering
	\caption{Mean value of systolic blood pressure (SBP), diastolic blood pressure (DBP), aortic pulse pressure ($PP=SBP-DBP$), mean blood pressure (MBP), and the maximum of the blood low (BF) at the level of of ascending aorta for 4,374  virtual subject based in-silico database.} 
\begin{tabular}{|c|c|c|c|c|c|c|c|}
\hline
Age                 & Number of subjects & Heart rate & SBP [mmHg]         & DBP [mmHg]        & MBP [mmHg]        & APP [mmHg]         & BF [ml/s]          \\ \hline \hline
\multirow{ }{ }{25} & 243                        & 84.11      & 101.07 $\pm$ 7.87  & 76.67 $\pm$ 5.98  & 90.75 $\pm$ 6.22  & 24.3957 $\pm$ 7.09 & 404.69 $\pm$ 53.87 \\ \cline{2-8} 
                    & 243                        & 72.91      & 99.95 $\pm$ 8.41   & 74.58 $\pm$ 5.44  & 89.13 $\pm$ 6.21  & 25.36 $\pm$ 7.29   & 389.58 $\pm$ 52.04 \\ \cline{2-8} 
                    & 243                        & 61.72      & 99.04 $\pm$ 8.93   & 72.54 $\pm$ 4.99  & 87.43 $\pm$ 6.31  & 26.49 $\pm$ 7.37   & 376.65 $\pm$ 50.93 \\ \hline \hline
\multirow{3}{*}{35} & 243                        & 87.99      & 104.549 $\pm$ 8.59 & 78.59 $\pm$ 6.24  & 93.60 $\pm$ 6.18  & 25.95 $\pm$ 8.96   & 394.46 $\pm$ 51.97 \\ \cline{2-8} 
                    & 243                        & 76.80      & 104.36 $\pm$ 8.94  & 77.33 $\pm$ 5.93  & 92.78 $\pm$ 6.19  & 27.02 $\pm$ 9.02   & 378.43 $\pm$ 50.56 \\ \cline{2-8} 
                    & 243                        & 65.60      & 104.07 $\pm$ 9.33  & 75.91 $\pm$ 5.50  & 91.58 $\pm$ 6.26  & 28.15 $\pm$ 9.00   & 363.64 $\pm$ 49.50 \\ \hline \hline
\multirow{3}{*}{45} & 243                        & 88.93      & 110.01 $\pm$ 9.09  & 79.53 $\pm$ 6.45  & 96.86 $\pm$ 6.16  & 30.47 $\pm$ 10.05  & 384.97 $\pm$ 50.42 \\ \cline{2-8} 
                    & 243                        & 77.74      & 110.15 $\pm$ 9.27  & 78.58 $\pm$ 6.18  & 96.22 $\pm$ 6.17  & 31.57 $\pm$ 9.97   & 368.01 $\pm$ 49.07 \\ \cline{2-8} 
                    & 243                        & 66.54      & 110.18 $\pm$ 9.67  & 77.39 $\pm$ 5.80  & 95.24 $\pm$ 6.24  & 32.78 $\pm$ 10.06  & 352.91 $\pm$ 48.21 \\ \hline \hline
\multirow{3}{*}{55} & 243                        & 88.49      & 112.50 $\pm$ 10.35 & 77.56 $\pm$ 6.64  & 96.96 $\pm$ 6.12  & 34.93 $\pm$ 12.37  & 374.06 $\pm$ 48.72 \\ \cline{2-8} 
                    & 243                        & 77.30      & 112.88 $\pm$ 10.55 & 76.76 $\pm$ 6.40  & 96.42 $\pm$ 6.15  & 36.11 $\pm$ 12.37  & 357.28 $\pm$ 47.57 \\ \cline{2-8} 
                    & 243                        & 66.10      & 113.15 $\pm$ 10.94 & 75.84 $\pm$ 6.05  & 95.62 $\pm$ 6.22  & 37.30 $\pm$ 12.47  & 341.64 $\pm$ 46.86 \\ \hline \hline
\multirow{3}{*}{65} & 243                        & 87.70      & 115.83 $\pm$ 21.00 & 75.46 $\pm$ 21.47 & 97.13 $\pm$ 19.69 & 41.96 $\pm$ 26.98  & 362.19 $\pm$ 46.88 \\ \cline{2-8} 
                    & 243                        & 76.51      & 116.26 $\pm$ 21.03 & 74.91 $\pm$ 21.43 & 96.74 $\pm$ 19.72 & 42.94 $\pm$ 26.85  & 345.21 $\pm$ 45.91 \\ \cline{2-8} 
                    & 243                        & 65.31      & 116.70 $\pm$ 21.20 & 74.21 $\pm$ 21.36 & 96.08 $\pm$ 19.78 & 44.08 $\pm$ 26.82  & 329.64 $\pm$ 45.47 \\ \hline \hline
\multirow{3}{*}{75} & 243                        & 85.63      & 117.42 $\pm$ 23.41 & 71.69 $\pm$ 22.14 & 95.34 $\pm$ 20.02 & 49.17 $\pm$ 29.85  & 341.50 $\pm$ 61.49 \\ \cline{2-8} 
                    & 243                        & 74.43      & 118.10 $\pm$ 14.39 & 70.02 $\pm$ 8.46  & 94.52 $\pm$ 6.02  & 48.07 $\pm$ 19.94  & 331.84 $\pm$ 44.34 \\ \cline{2-8} 
                    & 243                        & 63.23      & 118.31 $\pm$ 15.25 & 68.95 $\pm$ 7.73  & 93.56 $\pm$ 6.26  & 49.35 $\pm$ 20.08  & 316.91 $\pm$ 43.86 \\ \hline
\end{tabular}
\end{table}
\end{landscape}
\newpage
\begin{landscape}
\begin{table}[p]
\caption{Mean values of the goodness of fit criterion
($\mathrm {NRMSE}$, $\mathrm {Deviation [\%]}$, and $\mathrm {AIC_c}$ of each age and heart rate based-group.}
\begin{center}
	\centering
	\resizebox{1\linewidth}{!}{
		\renewcommand{\arraystretch}{3}
\begin{tabular}{cc|l|l|c|l|l|l|l|l|l|l|l|l|l|l|l|l|l|l|l|l|l|}
\cline{3-23}
 &  & \multicolumn{3}{c|}{Model A} & \multicolumn{3}{c|}{Model B} & \multicolumn{3}{c|}{Model C} & \multicolumn{3}{c|}{Model D} & \multicolumn{3}{c|}{Model E} & \multicolumn{3}{c|}{Model F} & \multicolumn{3}{c|}{Model G} \\ \cline{3-23} \hline
\multicolumn{1}{|c|}{Age} & Heart rate & RMSE & Deviation \% & $AIC_c$ & RMSE & Deviation \% & $AIC_c$ & RMSE & Deviation \% & $AIC_c$ & RMSE & Deviation \% & $AIC_c$ & RMSE & Deviation \% & $AIC_c$ & RMSE & Deviation \% & $AIC_c$ & RMSE & Deviation \% & $AIC_c$ \\ \hline
\multicolumn{1}{|c|}{\multirow{3}{*}{25}} & 84.11 & 0.42 $\pm$0.04 & 37.04 $\pm$4.88 & 6.74 $\pm$0.18 & 0.22 $\pm$0.01 & 17.75 $\pm$1.44 & 11.18 $\pm$0.12 & 0.4 $\pm$0.04 & 34.14 $\pm$4.18 & 6.82 $\pm$0.18 & 0.26 $\pm$0.01 & 21.14 $\pm$1.39 & 10.86 $\pm$0.11 & 0.18 $\pm$0.01 & 13.11 $\pm$1.07 & 15.47 $\pm$0.14 & 0.17 $\pm$0.02 & 12.76 $\pm$1.27 & 57.51 $\pm$0.18 & 0.29 $\pm$0.02 & 23.02 $\pm$1.72 & 7.47 $\pm$0.14 \\ \cline{2-23} 
\multicolumn{1}{|c|}{} & 72.91 & 0.45 $\pm$0.04 & 42.93 $\pm$5.09 & 6.45 $\pm$0.17 & 0.23 $\pm$0.01 & 17.9 $\pm$1.69 & 10.82 $\pm$0.12 & 0.43 $\pm$0.04 & 38.4 $\pm$4.05 & 6.53 $\pm$0.17 & 0.26 $\pm$0.01 & 21.13 $\pm$1.32 & 10.52 $\pm$0.1 & 0.17 $\pm$0.01 & 12.85 $\pm$0.93 & 14.84 $\pm$0.13 & 0.18 $\pm$0.02 & 12.73 $\pm$0.82 & 47.19 $\pm$0.17 & 0.3 $\pm$0.02 & 23.49 $\pm$1.67 & 7.27 $\pm$0.14 \\ \cline{2-23} 
\multicolumn{1}{|c|}{} & 61.72 & 0.49 $\pm$0.04 & 51.51 $\pm$5.99 & 6.13 $\pm$0.15 & 0.23 $\pm$0.01 & 18.12 $\pm$1.71 & 10.46 $\pm$0.12 & 0.47 $\pm$0.04 & 44.25 $\pm$4.4 & 6.22 $\pm$0.15 & 0.26 $\pm$0.01 & 21 $\pm$1.13 & 10.2 $\pm$0.1 & 0.17 $\pm$0.01 & 12.53 $\pm$0.71 & 14.2 $\pm$0.11 & 0.19 $\pm$0.02 & 12.98 $\pm$1.01 & 39.37 $\pm$0.21 & 0.31 $\pm$0.02 & 24.01 $\pm$1.5 & 7.05 $\pm$0.14 \\ \hline
\multicolumn{1}{|c|}{\multirow{3}{*}{35}} & 87.99 & 0.46 $\pm$0.04 & 39.65 $\pm$4.68 & 6.67 $\pm$0.2 & 0.24 $\pm$0.02 & 19.74 $\pm$2.11 & 11.22 $\pm$0.15 & 0.44 $\pm$0.04 & 36.78 $\pm$4.1 & 6.74 $\pm$0.2 & 0.29 $\pm$0.02 & 23.12 $\pm$1.82 & 10.9 $\pm$0.12 & 0.18 $\pm$0.01 & 13.96 $\pm$1.29 & 15.85 $\pm$0.16 & 0.18 $\pm$0.02 & 13.12 $\pm$1.67 & 66.49 $\pm$0.25 & 0.32 $\pm$0.02 & 25.17 $\pm$2.2 & 7.38 $\pm$0.14 \\ \cline{2-23} 
\multicolumn{1}{|c|}{} & 76.80 & 0.49 $\pm$0.04 & 45.01 $\pm$4.79 & 6.37 $\pm$0.17 & 0.25 $\pm$0.02 & 19.81 $\pm$2.26 & 10.8 $\pm$0.15 & 0.47 $\pm$0.04 & 40.6 $\pm$3.74 & 6.45 $\pm$0.17 & 0.29 $\pm$0.02 & 23.01 $\pm$1.65 & 10.5 $\pm$0.11 & 0.18 $\pm$0.01 & 13.82 $\pm$1.04 & 15.07 $\pm$0.15 & 0.18 $\pm$0.02 & 13.14 $\pm$1.34 & 51.43 $\pm$0.26 & 0.33 $\pm$0.02 & 25.53 $\pm$2.01 & 7.16 $\pm$0.14 \\ \cline{2-23} 
\multicolumn{1}{|c|}{} & 65.60 & 0.51 $\pm$0.04 & 51.49 $\pm$5.84 & 6.09 $\pm$0.15 & 0.25 $\pm$0.02 & 19.69 $\pm$2.4 & 10.4 $\pm$0.15 & 0.49 $\pm$0.04 & 44.93 $\pm$4.12 & 6.18 $\pm$0.15 & 0.28 $\pm$0.01 & 22.43 $\pm$1.54 & 10.15 $\pm$0.1 & 0.18 $\pm$0.01 & 13.3 $\pm$0.95 & 14.32 $\pm$0.14 & 0.19 $\pm$0.03 & 13.2 $\pm$1.5 & 41.37 $\pm$0.32 & 0.33 $\pm$0.02 & 25.4 $\pm$1.84 & 6.97 $\pm$0.13 \\ \hline
\multicolumn{1}{|c|}{\multirow{3}{*}{45}} & 88.93 & 0.51 $\pm$0.04 & 42.97 $\pm$4.51 & 6.46 $\pm$0.16 & 0.27 $\pm$0.02 & 21.49 $\pm$2.21 & 11.05 $\pm$0.14 & 0.49 $\pm$0.04 & 39.83 $\pm$3.69 & 6.53 $\pm$0.16 & 0.31 $\pm$0.02 & 24.49 $\pm$2.63 & 10.76 $\pm$0.15 & 0.19 $\pm$0.02 & 14.19 $\pm$1.48 & 15.83 $\pm$0.19 & 0.18 $\pm$0.03 & 12.8 $\pm$1.78 & 66.5 $\pm$0.3 & 0.35 $\pm$0.03 & 26.91 $\pm$2.65 & 7.22 $\pm$0.15 \\ \cline{2-23} 
\multicolumn{1}{|c|}{} & 77.74 & 0.53 $\pm$0.04 & 47.24 $\pm$5.72 & .19 $\pm$0.15 & 0.27 $\pm$0.02 & 21.35 $\pm$2.02 & 10.66 $\pm$0.13 & 0.51 $\pm$0.04 & 42.78 $\pm$4.31 & 6.27 $\pm$0.14 & 0.3 $\pm$0.02 & 24.08 $\pm$2.31 & 10.4 $\pm$0.15 & 0.18 $\pm$0.02 & 13.7 $\pm$1.3 & 15.04 $\pm$0.17 & 0.18 $\pm$0.03 & 12.91 $\pm$1.59 & 51.43 $\pm$0.33 & 0.35 $\pm$0.03 & 26.89 $\pm$2.49 & 7.04 $\pm$0.15 \\ \cline{2-23} 
\multicolumn{1}{|c|}{} & 66.54 & 0.56 $\pm$0.04 & 52.88 $\pm$7.67 & 5.93 $\pm$0.13 & 0.27 $\pm$0.02 & 21.38 $\pm$1.94 & 10.23 $\pm$0.12 & 0.53 $\pm$0.03 & 46.68 $\pm$5.58 & 6.02 $\pm$0.13 & 0.3 $\pm$0.02 & 23.63 $\pm$2.02 & 10.04 $\pm$0.13 & 0.18 $\pm$0.01 & 13.64 $\pm$1.09 & 14.28 $\pm$0.15 & 0.19 $\pm$0.04 & 12.91 $\pm$1.8 & 41.39 $\pm$0.38 & 0.35 $\pm$0.03 & 26.81 $\pm$2.35 & 6.85 $\pm$0.16 \\ \hline
\multicolumn{1}{|c|}{\multirow{3}{*}{55}} & 88.49 & 0.55 $\pm$0.04 & 43.46 $\pm$8.77 & 6.29 $\pm$0.15 & 0.27 $\pm$0.04 & 20.74 $\pm$4.86 & 11.07 $\pm$0.34 & 0.53 $\pm$0.04 & 40.12 $\pm$7.71 & 6.37 $\pm$0.15 & 0.31 $\pm$0.06 & 23.49 $\pm$6.13 & 10.81 $\pm$0.44 & 0.17 $\pm$0.03 & 12.97 $\pm$2.57 & 15.96 $\pm$0.34 & 0.18 $\pm$0.04 & 12.36 $\pm$2.32 & 66.51 $\pm$0.41 & 0.35 $\pm$0.06 & 26.09 $\pm$6.58 & 7.21 $\pm$0.39 \\ \cline{2-23} 
\multicolumn{1}{|c|}{} & \multicolumn{1}{l|}{77.3} & 0.57 $\pm$0.04 & 46.88 $\pm$10.37 & \multicolumn{1}{l|}{6.04 $\pm$0.14} & 0.27 $\pm$0.04 & 20.5 $\pm$4.76 & 10.67 $\pm$0.34 & 0.55 $\pm$0.04 & 42.48 $\pm$8.81 & 6.13 $\pm$0.14 & 0.3 $\pm$0.05 & 22.84 $\pm$5.82 & 10.46 $\pm$0.41 & 0.17 $\pm$0.02 & 12.67 $\pm$2.47 & 15.18 $\pm$0.31 & 0.18 $\pm$0.04 & 12.48 $\pm$2.42 & 51.47 $\pm$0.45 & 0.35 $\pm$0.06 & 25.79 $\pm$6.56 & 7.06 $\pm$0.42 \\ \cline{2-23} 
\multicolumn{1}{|c|}{} & \multicolumn{1}{l|}{66.1} & 0.59 $\pm$0.04 & 51.42 $\pm$12.43 & \multicolumn{1}{l|}{5.79 $\pm$0.13} & 0.28 $\pm$0.04 & 20.83 $\pm$4.75 & 10.21 $\pm$0.33 & 0.57 $\pm$0.04 & 45.74 $\pm$10.26 & 5.9 $\pm$0.13 & 0.3 $\pm$0.05 & 22.72 $\pm$5.42 & 10.05 $\pm$0.37 & 0.17 $\pm$0.02 & 12.57 $\pm$2.3 & 14.38 $\pm$0.27 & 0.18 $\pm$0.05 & 12.41 $\pm$2.4 & 41.45 $\pm$0.49 & 0.35 $\pm$0.06 & 25.83 $\pm$6.52 & 6.88 $\pm$0.43 \\ \hline
\multicolumn{1}{|l|}{\multirow{3}{*}{65}} & \multicolumn{1}{l|}{87.7} & 0.59 $\pm$0.04 & 41.16 $\pm$12.44 & \multicolumn{1}{l|}{6.14 $\pm$0.14} & 0.26 $\pm$0.06 & 18.33 $\pm$6.92 & 11.19 $\pm$0.54 & 0.56 $\pm$0.04 & 37.93 $\pm$10.96 & 6.24 $\pm$0.15 & 0.29 $\pm$0.08 & 20.59 $\pm$8.5 & 10.97 $\pm$0.64 & 0.16 $\pm$0.03 & 11.21 $\pm$3.45 & 16.2 $\pm$0.47 & 0.18 $\pm$0.05 & 12.34 $\pm$3.7 & 66.54 $\pm$0.52 & 0.34 $\pm$0.09 & 23.23 $\pm$9.38 & 7.33 $\pm$0.62 \\ \cline{2-23} 
\multicolumn{1}{|l|}{} & \multicolumn{1}{l|}{76.51} & 0.62 $\pm$0.04 & 43.99 $\pm$14.09 & \multicolumn{1}{l|}{5.89 $\pm$0.13} & 0.26 $\pm$0.06 & 18.37 $\pm$7.07 & 10.74 $\pm$0.55 & 0.59 $\pm$0.04 & 40.15 $\pm$12.26 & 6 $\pm$0.15 & 0.29 $\pm$0.08 & 20.32 $\pm$8.32 & 10.57 $\pm$0.62 & 0.16 $\pm$0.04 & 11.26 $\pm$3.62 & 15.34 $\pm$0.47 & 0.18 $\pm$0.05 & 12.58 $\pm$3.93 & 51.46 $\pm$0.57 & 0.34 $\pm$0.09 & 23.24 $\pm$9.54 & 7.15 $\pm$0.65 \\ \cline{2-23} 
\multicolumn{1}{|l|}{} & \multicolumn{1}{l|}{65.31} & 0.63 $\pm$0.04 & 47.08 $\pm$15.83 & \multicolumn{1}{l|}{5.67 $\pm$0.12} & 0.27 $\pm$0.06 & 18.72 $\pm$7.09 & 10.28 $\pm$0.55 & 0.6 $\pm$0.04 & 42.45 $\pm$13.52 & 5.79 $\pm$0.15 & 0.29 $\pm$0.07 & 20.3 $\pm$7.7 & 10.13 $\pm$0.54 & 0.16 $\pm$0.03 & 10.94 $\pm$3.29 & 14.56 $\pm$0.38 & 0.18 $\pm$0.05 & 12.16 $\pm$3.51 & 41.52 $\pm$0.56 & 0.34 $\pm$0.09 & 23.09 $\pm$9.38 & 7.01 $\pm$0.67 \\ \hline
\multicolumn{1}{|l|}{\multirow{3}{*}{75}} & \multicolumn{1}{l|}{85.63} & 0.63 $\pm$0.04 & 42.07 $\pm$15.6 & \multicolumn{1}{l|}{5.93 $\pm$0.13} & 0.27 $\pm$0.07 & 17.71 $\pm$7.77 & 10.89 $\pm$0.58 & 0.6 $\pm$0.04 & 38.6 $\pm$13.58 & 6.03 $\pm$0.14 & 0.3 $\pm$0.09 & 19.8 $\pm$9.33 & 10.72 $\pm$0.66 & 0.16 $\pm$0.04 & 10.6 $\pm$3.84 & 15.73 $\pm$0.48 & 0.19 $\pm$0.06 & 12.97 $\pm$4.96 & 57.41 $\pm$0.65 & 0.35 $\pm$0.1 & 22.47 $\pm$10.47 & 7.24 $\pm$0.7 \\ \cline{2-23} 
\multicolumn{1}{|l|}{} & \multicolumn{1}{l|}{74.43} & 0.66 $\pm$0.04 & 43.99 $\pm$17.53 & \multicolumn{1}{l|}{5.7 $\pm$0.13} & 0.27 $\pm$0.07 & 17.09 $\pm$7.73 & 10.56 $\pm$0.6 & 0.62 $\pm$0.04 & 40.3 $\pm$15.3 & 5.82 $\pm$0.14 & 0.29 $\pm$0.08 & 18.83 $\pm$8.86 & 10.42 $\pm$0.66 & 0.16 $\pm$0.03 & 10.17 $\pm$3.78 & 15.1 $\pm$0.46 & 0.2 $\pm$0.07 & 12.94 $\pm$5.08 & 47.08 $\pm$0.68 & 0.34 $\pm$0.11 & 21.69 $\pm$10.42 & 7.13 $\pm$0.74 \\ \cline{2-23} 
\multicolumn{1}{|l|}{} & \multicolumn{1}{l|}{63.23} & 0.68 $\pm$0.04 & 46.48 $\pm$19.93 & \multicolumn{1}{l|}{5.49 $\pm$0.13} & 0.27 $\pm$0.07 & 17.23 $\pm$7.96 & 10.17 $\pm$0.61 & 0.63 $\pm$0.05 & 42.31 $\pm$17.29 & 5.62 $\pm$0.15 & 0.29 $\pm$0.08 & 18.3 $\pm$8.77 & 10.09 $\pm$0.65 & 0.16 $\pm$0.03 & 9.85 $\pm$3.61 & 14.4 $\pm$0.42 & 0.2 $\pm$0.07 & 12.85 $\pm$5.17 & 39.36 $\pm$0.69 & 0.34 $\pm$0.11 & 21.39 $\pm$10.46 & 7 $\pm$0.76 \\ \hline
\end{tabular}
}
\end{center}
\end{table}
\end{landscape}
\begin{landscape}
\begin{table}[p]
\caption{Mean value of the parameter estimates of the fractional-order models for each age and heart rate based-group.}
\begin{center}
	\centering
	\resizebox{1\linewidth}{!}{
		\renewcommand{\arraystretch}{2}
\begin{tabular}{cc|c|c|c|c|c|c|c|c|c|c|c|c|c|l|}
\cline{3-16}
 &  & \multicolumn{2}{c|}{Model A} & \multicolumn{3}{c|}{Model B} & \multicolumn{2}{c|}{Model C} & \multicolumn{3}{c|}{Model D} & \multicolumn{4}{c|}{Model E} \\ \hline
\multicolumn{1}{|c|}{Age} & Heart rate & $\mathrm{C_{\alpha_A}}$ & $\mathrm{\alpha_A}$ & $\mathrm{R_B}$ & $\mathrm{C_{\alpha_B}}$ & $\mathrm{\alpha_B}$ & $\mathrm{C_{stat_C}, \ C_{\alpha_C}}$ & $\mathrm{\alpha_C}$ & $\mathrm{R_D}$ & $\mathrm{C_{stat_D}}, \ \mathrm{C_{\alpha_D}}$ & $\mathrm{\alpha_D}$ & $\mathrm{R_{1_E}}$ & $\mathrm{R_{2_E}}$ & $\mathrm{C_{\alpha_E}}$ & $\mathrm{\alpha_E}$ \\ \hline
\multicolumn{1}{|c|}{\multirow{3}{*}{25}} & 84.11 & 4.44 $\pm$ 1.32 & 0.38 $\pm$ 0.08 & 0.05 $\pm$0.01 & 0.87 $\pm$0.11 & 1.33 $\pm$0.04 & 6 $\pm$1.52 & 0.31 $\pm$0.06 & 0.05 $\pm$0.01 & 2.27 $\pm$0.46 & 1.44 $\pm$0.01 & 0.74 $\pm$0.12 & 37.91 $\pm$ 40.48 & 0.08 $\pm$ 0.12 & 1.29 $\pm$ 0.04 \\ \cline{2-16} 
\multicolumn{1}{|c|}{} & 72.91 & 3.23 $\pm$0.76 & 0.47 $\pm$0.07 & 0.05 $\pm$0.01 & 0.99 $\pm$0.14 & 1.29 $\pm$0.04 & 4.74 $\pm$0.97 & 0.38 $\pm$0.06 & 0.05 $\pm$0.01 & 2.19 $\pm$0.39 & 1.46 $\pm$0.01 & 0.72 $\pm$0.12 & 43.34 $\pm$45.96 & 0.06 $\pm$0.09 & 1.28 $\pm$0.03 \\ \cline{2-16} 
\multicolumn{1}{|c|}{} & 61.72 & 2.37 $\pm$0.37 & 0.56 $\pm$0.06 & 0.05 $\pm$0.01 & 1.12 $\pm$0.17 & 1.24 $\pm$0.03 & 3.78 $\pm$0.55 & 0.45 $\pm$0.05 & 0.05 $\pm$0.01 & 2.17 $\pm$0.33 & 1.48 $\pm$0.02 & 0.72 $\pm$0.11 & 36.76 $\pm$43.1 & 0.06 $\pm$0.08 & 1.28 $\pm$0.03 \\ \hline
\multicolumn{1}{|c|}{\multirow{3}{*}{35}} & 87.99 & 3.76 $\pm$1.37 & 0.42 $\pm$0.08 & 0.05 $\pm$0.01 & 0.77 $\pm$0.11 & 1.33 $\pm$0.06 & 5.2 $\pm$1.63 & 0.34 $\pm$0.07 & 0.05 $\pm$0.01 & 2.01 $\pm$0.5 & 1.44 $\pm$0.01 & 0.83 $\pm$0.18 & 61.39 $\pm$61.54 & 0.06 $\pm$0.1 & 1.31 $\pm$0.04 \\ \cline{2-16} 
\multicolumn{1}{|c|}{} & 76.80 & 2.78 $\pm$0.78 & 0.5 $\pm$0.07 & 0.05 $\pm$0.01 & 0.87 $\pm$0.14 & 1.29 $\pm$0.05 & 4.16 $\pm$1.03 & 0.41 $\pm$0.06 & 0.05 $\pm$0.01 & 1.95 $\pm$0.42 & 1.45 $\pm$0.01 & 0.81 $\pm$0.18 & 59.49 $\pm$60.43 & 0.04 $\pm$0.08 & 1.3 $\pm$0.03 \\ \cline{2-16} 
\multicolumn{1}{|c|}{} & 65.60 & 2.18 $\pm$0.4 & 0.58 $\pm$0.05 & 0.05 $\pm$0.01 & 1 $\pm$0.18 & 1.24 $\pm$0.04 & 3.48 $\pm$0.62 & 0.46 $\pm$0.05 & 0.05 $\pm$0.01 & 1.97 $\pm$0.36 & 1.47 $\pm$0.03 & 0.79 $\pm$0.16 & 58.52 $\pm$64.54 & 0.05 $\pm$0.07 & 1.3 $\pm$0.03 \\ \hline
\multicolumn{1}{|c|}{\multirow{3}{*}{45}} & 88.93 & 2.56 $\pm$0.74 & 0.5 $\pm$0.07 & 0.06 $\pm$0.01 & 0.68 $\pm$0.1 & 1.29 $\pm$0.06 & 3.76 $\pm$0.94 & 0.41 $\pm$0.06 & 0.05 $\pm$0.01 & 1.62 $\pm$0.34 & 1.43 $\pm$0.01 & 0.96 $\pm$0.23 & 87.44 $\pm$63.63 & 0.01 $\pm$0.04 & 1.32 $\pm$0.03 \\ \cline{2-16} 
\multicolumn{1}{|c|}{} & 77.74 & 2.03 $\pm$0.42 & 0.56 $\pm$0.06 & 0.06 $\pm$0.01 & 0.77 $\pm$0.12 & 1.25 $\pm$0.06 & 3.18 $\pm$0.61 & 0.46 $\pm$0.05 & 0.05 $\pm$0.01 & 1.61 $\pm$0.3 & 1.44 $\pm$0.02 & 0.95 $\pm$0.22 & 84.43 $\pm$66.85 & 0.01 $\pm$0.03 & 1.31 $\pm$0.04 \\ \cline{2-16} 
\multicolumn{1}{|c|}{} & 66.54 & 1.7 $\pm$0.24 & 0.62 $\pm$0.05 & 0.05 $\pm$0.01 & 0.87 $\pm$0.15 & 1.21 $\pm$0.05 & 2.79 $\pm$0.4 & 0.51 $\pm$0.05 & 0.05 $\pm$0.01 & 1.66 $\pm$0.26 & 1.44 $\pm$0.05 & 0.93 $\pm$0.21 & 94.17 $\pm$79.48 & 0.01 $\pm$0.03 & 1.31 $\pm$0.04 \\ \hline
\multicolumn{1}{|c|}{\multirow{3}{*}{55}} & 88.49 & 1.88 $\pm$0.47 & 0.57 $\pm$0.07 & 0.06 $\pm$0.01 & 0.64 $\pm$0.09 & 1.25 $\pm$0.08 & 2.91 $\pm$0.64 & 0.47 $\pm$0.06 & 0.05 $\pm$0.01 & 1.39 $\pm$0.26 & 1.41 $\pm$0.06 & 1.1 $\pm$0.29 & 101.15 $\pm$68.16 & 0.01 $\pm$0.02 & 1.32 $\pm$0.06 \\ \cline{2-16} 
\multicolumn{1}{|c|}{} & \multicolumn{1}{l|}{77.30} & \multicolumn{1}{l|}{1.59 $\pm$0.28} & \multicolumn{1}{l|}{0.62 $\pm$0.05} & \multicolumn{1}{l|}{0.05 $\pm$0.01} & \multicolumn{1}{l|}{0.72 $\pm$0.11} & \multicolumn{1}{l|}{1.21 $\pm$0.07} & \multicolumn{1}{l|}{2.6 $\pm$0.43} & \multicolumn{1}{l|}{0.52 $\pm$0.05} & \multicolumn{1}{l|}{0.05 $\pm$0.01} & \multicolumn{1}{l|}{1.42 $\pm$0.25} & \multicolumn{1}{l|}{1.42 $\pm$0.11} & \multicolumn{1}{l|}{1.08 $\pm$0.27} & \multicolumn{1}{l|}{106.07 $\pm$68.67} & \multicolumn{1}{l|}{0.01 $\pm$0.02} & 1.31 $\pm$0.06 \\ \cline{2-16} 
\multicolumn{1}{|c|}{} & \multicolumn{1}{l|}{66.10} & \multicolumn{1}{l|}{1.41 $\pm$0.18} & \multicolumn{1}{l|}{0.66 $\pm$0.04} & \multicolumn{1}{l|}{0.05 $\pm$0.01} & \multicolumn{1}{l|}{0.8 $\pm$0.13} & \multicolumn{1}{l|}{1.17 $\pm$0.06} & \multicolumn{1}{l|}{2.38 $\pm$0.31} & \multicolumn{1}{l|}{0.55 $\pm$0.04} & \multicolumn{1}{l|}{0.05 $\pm$0.01} & \multicolumn{1}{l|}{1.48 $\pm$0.26} & \multicolumn{1}{l|}{1.44 $\pm$0.17} & \multicolumn{1}{l|}{1.05 $\pm$0.27} & \multicolumn{1}{l|}{116.23 $\pm$74.93} & \multicolumn{1}{l|}{0.01 $\pm$0.02} & 1.3 $\pm$0.06 \\ \hline
\multicolumn{1}{|l|}{\multirow{3}{*}{65}} & \multicolumn{1}{l|}{87.70} & \multicolumn{1}{l|}{1.46 $\pm$0.31} & \multicolumn{1}{l|}{0.63 $\pm$0.05} & \multicolumn{1}{l|}{0.05 $\pm$0.01} & \multicolumn{1}{l|}{0.61 $\pm$0.09} & \multicolumn{1}{l|}{1.2 $\pm$0.09} & \multicolumn{1}{l|}{2.38 $\pm$0.46} & \multicolumn{1}{l|}{0.53 $\pm$0.05} & \multicolumn{1}{l|}{0.05 $\pm$0.01} & \multicolumn{1}{l|}{1.24 $\pm$0.22} & \multicolumn{1}{l|}{1.39 $\pm$0.13} & \multicolumn{1}{l|}{1.27 $\pm$0.36} & \multicolumn{1}{l|}{98.57 $\pm$58.23} & \multicolumn{1}{l|}{0 $\pm$0.01} & 1.31 $\pm$0.08 \\ \cline{2-16} 
\multicolumn{1}{|l|}{} & \multicolumn{1}{l|}{76.51} & \multicolumn{1}{l|}{1.29 $\pm$0.2} & \multicolumn{1}{l|}{0.67 $\pm$0.04} & \multicolumn{1}{l|}{0.05 $\pm$0.01} & \multicolumn{1}{l|}{0.67 $\pm$0.11} & \multicolumn{1}{l|}{1.17 $\pm$0.07} & \multicolumn{1}{l|}{2.18 $\pm$0.33} & \multicolumn{1}{l|}{0.57 $\pm$0.04} & \multicolumn{1}{l|}{0.05 $\pm$0.01} & \multicolumn{1}{l|}{1.26 $\pm$0.24} & \multicolumn{1}{l|}{1.42 $\pm$0.2} & \multicolumn{1}{l|}{1.25 $\pm$0.34} & \multicolumn{1}{l|}{103.91 $\pm$61.15} & \multicolumn{1}{l|}{0 $\pm$0.01} & 1.30 $\pm$0.08 \\ \cline{2-16} 
\multicolumn{1}{|l|}{} & \multicolumn{1}{l|}{65.31} & \multicolumn{1}{l|}{1.21 $\pm$0.15} & \multicolumn{1}{l|}{0.69 $\pm$0.03} & \multicolumn{1}{l|}{0.05 $\pm$0.01} & \multicolumn{1}{l|}{0.74 $\pm$0.13} & \multicolumn{1}{l|}{1.14 $\pm$0.06} & \multicolumn{1}{l|}{2.1 $\pm$0.27} & \multicolumn{1}{l|}{0.59 $\pm$0.04} & \multicolumn{1}{l|}{0.05 $\pm$0.01} & \multicolumn{1}{l|}{1.3 $\pm$0.3} & \multicolumn{1}{l|}{1.48 $\pm$0.28} & \multicolumn{1}{l|}{1.21 $\pm$0.33} & \multicolumn{1}{l|}{133.47 $\pm$69.49} & \multicolumn{1}{l|}{0 $\pm$0.01} & 1.29 $\pm$0.08 \\ \hline
\multicolumn{1}{|l|}{\multirow{3}{*}{75}} & \multicolumn{1}{l|}{85.63} & \multicolumn{1}{l|}{1.15 $\pm$0.26} & \multicolumn{1}{l|}{0.68 $\pm$0.04} & \multicolumn{1}{l|}{0.06 $\pm$0.01} & \multicolumn{1}{l|}{0.56 $\pm$0.1} & \multicolumn{1}{l|}{1.17 $\pm$0.09} & \multicolumn{1}{l|}{1.94 $\pm$0.41} & \multicolumn{1}{l|}{0.58 $\pm$0.04} & \multicolumn{1}{l|}{0.05 $\pm$0.01} & \multicolumn{1}{l|}{1.08 $\pm$0.24} & \multicolumn{1}{l|}{1.39 $\pm$0.21} & \multicolumn{1}{l|}{1.5 $\pm$0.51} & \multicolumn{1}{l|}{115.38 $\pm$71.2} & \multicolumn{1}{l|}{0 $\pm$0.01} & 1.31 $\pm$0.09 \\ \cline{2-16} 
\multicolumn{1}{|l|}{} & \multicolumn{1}{l|}{74.43} & \multicolumn{1}{l|}{1.04 $\pm$0.2} & \multicolumn{1}{l|}{0.71 $\pm$0.03} & \multicolumn{1}{l|}{0.05 $\pm$0.01} & \multicolumn{1}{l|}{0.62 $\pm$0.12} & \multicolumn{1}{l|}{1.14 $\pm$0.07} & \multicolumn{1}{l|}{1.83 $\pm$0.33} & \multicolumn{1}{l|}{0.61 $\pm$0.03} & \multicolumn{1}{l|}{0.05 $\pm$0.01} & \multicolumn{1}{l|}{1.1 $\pm$0.26} & \multicolumn{1}{l|}{1.44 $\pm$0.28} & \multicolumn{1}{l|}{1.47 $\pm$0.5} & \multicolumn{1}{l|}{119.61 $\pm$79.96} & \multicolumn{1}{l|}{0 $\pm$0.01} & 1.3 $\pm$0.09 \\ \cline{2-16} 
\multicolumn{1}{|l|}{} & \multicolumn{1}{l|}{63.23} & \multicolumn{1}{l|}{0.97 $\pm$0.16} & \multicolumn{1}{l|}{0.74 $\pm$0.04} & \multicolumn{1}{l|}{0.05 $\pm$0.01} & \multicolumn{1}{l|}{0.66 $\pm$0.14} & \multicolumn{1}{l|}{1.11 $\pm$0.06} & \multicolumn{1}{l|}{1.75 $\pm$0.28} & \multicolumn{1}{l|}{0.63 $\pm$0.04} & \multicolumn{1}{l|}{0.05 $\pm$0.01} & \multicolumn{1}{l|}{1.22 $\pm$0.24} & \multicolumn{1}{l|}{1.31 $\pm$0.23} & \multicolumn{1}{l|}{1.44 $\pm$0.48} & \multicolumn{1}{l|}{138.33 $\pm$80.26} & \multicolumn{1}{l|}{0 $\pm$0.01} & 1.29 $\pm$0.1 \\ \hline
\end{tabular}
}
\end{center}
\end{table}
\end{landscape}
\bibliographystyle{IEEEtran}
\bibliography{bibfile}

\begin{thebibliography}{10}
\providecommand{\url}[1]{#1}
\csname url@samestyle\endcsname
\providecommand{\newblock}{\relax}
\providecommand{\bibinfo}[2]{#2}
\providecommand{\BIBentrySTDinterwordspacing}{\spaceskip=0pt\relax}
\providecommand{\BIBentryALTinterwordstretchfactor}{4}
\providecommand{\BIBentryALTinterwordspacing}{\spaceskip=\fontdimen2\font plus
\BIBentryALTinterwordstretchfactor\fontdimen3\font minus
  \fontdimen4\font\relax}
\providecommand{\BIBforeignlanguage}[2]{{%
\expandafter\ifx\csname l@#1\endcsname\relax
\typeout{** WARNING: IEEEtran.bst: No hyphenation pattern has been}%
\typeout{** loaded for the language `#1'. Using the pattern for}%
\typeout{** the default language instead.}%
\else
\language=\csname l@#1\endcsname
\fi
#2}}
\providecommand{\BIBdecl}{\relax}
\BIBdecl

\bibitem{liang2005closed}
F.~Liang and H.~Liu, ``A closed-loop lumped parameter computational model for
  human cardiovascular system,'' \emph{JSME International Journal Series C
  Mechanical Systems, Machine Elements and Manufacturing}, vol.~48, no.~4, pp.
  484--493, 2005.

\bibitem{alderliesten2004simulation}
T.~Alderliesten, M.~K. Konings, and W.~J. Niessen, ``Simulation of minimally
  invasive vascular interventions for training purposes,'' \emph{Computer Aided
  Surgery}, vol.~9, no. 1-2, pp. 3--15, 2004.

\bibitem{van2016patient}
E.~Van~Disseldorp, N.~Petterson, M.~Rutten, F.~Van De~Vosse, M.~van Sambeek,
  and R.~Lopata, ``Patient specific wall stress analysis and mechanical
  characterization of abdominal aortic aneurysms using 4d ultrasound,''
  \emph{European Journal of Vascular and Endovascular Surgery}, vol.~52, no.~5,
  pp. 635--642, 2016.

\bibitem{beulen2011toward}
B.~W. Beulen, N.~Bijnens, G.~G. Koutsouridis, P.~J. Brands, M.~C. Rutten, and
  F.~N. van~de Vosse, ``Toward noninvasive blood pressure assessment in
  arteries by using ultrasound,'' \emph{Ultrasound in medicine \& biology},
  vol.~37, no.~5, pp. 788--797, 2011.

\bibitem{huberts2018needed}
W.~Huberts, S.~G. Heinen, N.~Zonnebeld, D.~A. van~den Heuvel, J.-P.~P.
  de~Vries, J.~H. Tordoir, D.~R. Hose, T.~Delhaas, and F.~N. van~de Vosse,
  ``What is needed to make cardiovascular models suitable for clinical decision
  support? a viewpoint paper,'' \emph{Journal of computational science},
  vol.~24, pp. 68--84, 2018.

\bibitem{leguy2010estimation}
C.~Leguy, E.~Bosboom, H.~Gelderblom, A.~Hoeks, and F.~Van De~Vosse,
  ``Estimation of distributed arterial mechanical properties using a wave
  propagation model in a reverse way,'' \emph{Medical engineering \& physics},
  vol.~32, no.~9, pp. 957--967, 2010.

\bibitem{stergiopulos1995evaluation}
N.~Stergiopulos, J.~Meister, and N.~Westerhof, ``Evaluation of methods for
  estimation of total arterial compliance,'' \emph{American Journal of
  Physiology-Heart and Circulatory Physiology}, vol. 268, no.~4, pp.
  H1540--H1548, 1995.

\bibitem{stergiopulos1999total}
N.~Stergiopulos, B.~E. Westerhof, and N.~Westerhof, ``Total arterial inertance
  as the fourth element of the windkessel model,'' \emph{American Journal of
  Physiology-Heart and Circulatory Physiology}, vol. 276, no.~1, pp. H81--H88,
  1999.

\bibitem{quick1998apparent}
C.~M. Quick, D.~S. Berger, and A.~Noordergraaf, ``Apparent arterial
  compliance,'' \emph{American Journal of Physiology-Heart and Circulatory
  Physiology}, vol. 274, no.~4, pp. H1393--H1403, 1998.

\bibitem{quick2000true}
C.~M. Quick, D.~S. Berger, D.~A. Hettrick, and A.~Noordergraaf, ``True arterial
  system compliance estimated from apparent arterial compliance,'' \emph{Annals
  of biomedical engineering}, vol.~28, no.~3, pp. 291--301, 2000.

\bibitem{craiem2003new}
D.~Craiem and R.~Armentano, ``The new apparent compliance concept as a simple
  lumped model,'' \emph{Cardiovascular Engineering: An International Journal},
  vol.~3, no.~2, pp. 81--83, 2003.

\bibitem{burattini1998complex}
R.~Burattini and S.~Natalucci, ``Complex and frequency-dependent compliance of
  viscoelastic windkessel resolves contradictions in elastic windkessels,''
  \emph{Medical engineering \& physics}, vol.~20, no.~7, pp. 502--514, 1998.

\bibitem{visaria2005modeling}
R.~Visaria, \emph{Modeling of cardiovascular system, pulmonary mechanics and
  gas exchange}.\hskip 1em plus 0.5em minus 0.4em\relax The University of Utah,
  2005.

\bibitem{ionescu2010modeling}
C.~M. Ionescu, J.~T. Machado, and R.~De~Keyser, ``Modeling of the lung
  impedance using a fractional-order ladder network with constant phase
  elements,'' \emph{IEEE Transactions on biomedical circuits and systems},
  vol.~5, no.~1, pp. 83--89, 2010.

\bibitem{kobayashi2012modeling}
Y.~Kobayashi, A.~Kato, H.~Watanabe, T.~Hoshi, K.~Kawamura, and M.~G. Fujie,
  ``Modeling of viscoelastic and nonlinear material properties of liver tissue
  using fractional calculations,'' \emph{Journal of Biomechanical Science and
  Engineering}, vol.~7, no.~2, pp. 177--187, 2012.

\bibitem{magin2006fractional}
R.~L. Magin, \emph{Fractional calculus in bioengineering}.\hskip 1em plus 0.5em
  minus 0.4em\relax Begell House Redding, 2006.

\bibitem{jaishankar2013power}
A.~Jaishankar and G.~H. McKinley, ``Power-law rheology in the bulk and at the
  interface: quasi-properties and fractional constitutive equations,''
  \emph{Proceedings of the Royal Society A: Mathematical, Physical and
  Engineering Sciences}, vol. 469, no. 2149, p. 20120284, 2013.

\bibitem{craiem2007fractional}
D.~Craiem and R.~L. Armentano, ``A fractional derivative model to describe
  arterial viscoelasticity,'' \emph{Biorheology}, vol.~44, no.~4, pp. 251--263,
  2007.

\bibitem{craiem2008fractional}
D.~Craiem, F.~J. Rojo, J.~M. Atienza, R.~L. Armentano, and G.~V. Guinea,
  ``Fractional-order viscoelasticity applied to describe uniaxial stress
  relaxation of human arteries,'' \emph{Physics in Medicine \& Biology},
  vol.~53, no.~17, p. 4543, 2008.

\bibitem{craiem2010fractional}
D.~Craiem and R.~L. Magin, ``Fractional order models of viscoelasticity as an
  alternative in the analysis of red blood cell (rbc) membrane mechanics,''
  \emph{Physical biology}, vol.~7, no.~1, p. 013001, 2010.

\bibitem{perdikaris2014fractional}
P.~Perdikaris and G.~E. Karniadakis, ``Fractional-order viscoelasticity in
  one-dimensional blood flow models,'' \emph{Annals of biomedical engineering},
  vol.~42, no.~5, pp. 1012--1023, 2014.

\bibitem{zerpa2015modeling}
J.~P. Zerpa, A.~Canelas, B.~Sensale, D.~B. Santana, and R.~Armentano,
  ``Modeling the arterial wall mechanics using a novel high-order viscoelastic
  fractional element,'' \emph{Applied Mathematical Modelling}, vol.~39, no.~16,
  pp. 4767--4780, 2015.

\bibitem{bahloul2018three}
M.~A. Bahloul and T.~M. Laleg-Kirati, ``Three-element fractional-order
  viscoelastic arterial windkessel model,'' in \emph{2018 40th Annual
  International Conference of the IEEE Engineering in Medicine and Biology
  Society (EMBC)}.\hskip 1em plus 0.5em minus 0.4em\relax IEEE, 2018, pp.
  5261--5266.

\bibitem{bahloul2018arterial}
------, ``Arterial viscoelastic model using lumped parameter circuit with
  fractional-order capacitor,'' in \emph{2018 IEEE 61st International Midwest
  Symposium on Circuits and Systems (MWSCAS)}.\hskip 1em plus 0.5em minus
  0.4em\relax IEEE, 2018, pp. 53--56.

\bibitem{bahloul2019fractional}
M.~A. Bahloul and T.-M.~L. Kirati, ``Fractional order models of arterial
  windkessel as an alternative in the analysis of the left ventricular
  afterload,'' \emph{arXiv preprint arXiv:1908.05239}, 2019.

\bibitem{nakagawa1992basic}
M.~Nakagawa and K.~Sorimachi, ``Basic characteristics of a fractance device,''
  \emph{IEICE Transactions on Fundamentals of Electronics, Communications and
  Computer Sciences}, vol.~75, no.~12, pp. 1814--1819, 1992.

\bibitem{tsirimokou2017systematic}
G.~Tsirimokou, ``A systematic procedure for deriving rc networks of
  fractional-order elements emulators using matlab,'' \emph{AEU-International
  Journal of Electronics and Communications}, vol.~78, pp. 7--14, 2017.

\bibitem{ionescu2013human}
C.~M. Ionescu, \emph{The human respiratory system: an analysis of the interplay
  between anatomy, structure, breathing and fractal dynamics}.\hskip 1em plus
  0.5em minus 0.4em\relax Springer Science \& Business Media, 2013.

\bibitem{doehring2005fractional}
T.~C. Doehring, A.~D. Freed, E.~O. Carew, and I.~Vesely, ``Fractional order
  viscoelasticity of the aortic valve cusp: an alternative to quasilinear
  viscoelasticity,'' \emph{Journal of biomechanical engineering}, vol. 127,
  no.~4, pp. 700--708, 2005.

\bibitem{charlton2019modelling}
P.~Charlton, J.~Mariscal~Harana, S.~Vennin, Y.~Li, P.~Chowienczyk, and
  J.~Alastruey, ``Modelling arterial pulse waves in healthy ageing: a database
  for in silico evaluation of haemodynamics and pulse wave indices,''
  \emph{American Journal of Physiology-Heart and Circulatory Physiology}, 2019.

\bibitem{coleman1996interior}
T.~F. Coleman and Y.~Li, ``An interior trust region approach for nonlinear
  minimization subject to bounds,'' \emph{SIAM Journal on optimization},
  vol.~6, no.~2, pp. 418--445, 1996.

\bibitem{goedhard1973model}
W.~Goedhard and A.~Knoop, ``A model of the arterial wall,'' \emph{Journal of
  biomechanics}, vol.~6, no.~3, pp. 281--288, 1973.

\end{thebibliography}

\end{document}